\documentclass[a4paper,fleqn]{cas-dc}
\usepackage[numbers]{natbib}
\usepackage{algorithmic}
\usepackage{array}
\usepackage[caption=false,font=normalsize,labelfont=sf,textfont=sf]{subfig}
\usepackage{textcomp}
\usepackage{stfloats}
\usepackage{url}
\usepackage{verbatim}
\usepackage{diagbox}
\usepackage{float}
\usepackage{multirow}
\usepackage{bbm}
\usepackage{amsfonts,graphicx,amsmath,amssymb,subfig,amsthm}
\usepackage{changepage}
\usepackage[most]{tcolorbox}
\usepackage{array}
\tcbset{enhanced, breakable}
\usepackage{caption}
\usepackage{enumitem}
\usepackage{xcolor}
\usepackage{subcaption}
\definecolor{deepgreen}{rgb}{0, 0.6, 0.2}
\hypersetup{
  hidelinks,
  colorlinks=true,
  linkcolor=blue,
  citecolor=deepgreen
}
\def\BibTeX{{\rm B\kern-.05em{\sc i\kern-.025em b}\kern-.08em
    T\kern-.1667em\lower.7ex\hbox{E}\kern-.125emX}}
\usepackage{balance}
\allowdisplaybreaks[4] 
\newtheorem{definition}{Definition} 
\newtheorem{theorem}{Theorem}

\begin{document}
\let\WriteBookmarks\relax
\def\floatpagepagefraction{1}
\def\textpagefraction{.001}
\shorttitle{Authenticated Private Set Intersection}
\shortauthors{Zixian Gong et~al.}
\title [mode = title]{Authenticated Private Set Intersection: A Merkle Tree-Based Approach for Enhancing Data Integrity}                      

\author[1]{Zixian Gong}[orcid={0009-0005-7059-5040}]
\ead{gzx@ruc.edu.cn}
\author[1,2,3]{Zhiyong Zheng}
\author[1]{Zhe Hu}
\author[1]{Kun Tian}
\author[1]{Yi Zhang}
\author[1]{Zhedanov Oleksiy}
\author[3]{Fengxia Liu}\corref{cor1}
\ead{shunliliu@gbu.edu.cn}

\affiliation[1]{organization={School of Mathematics, Renmin University of China},
                city={Beijing},
                postcode={100872}, 
                country={P. R. China}}

\affiliation[2]{organization={Institute of Mathematics, Henan Academy of Science},
                city={Zhengzhou},
                postcode={450046}, 
                state={Henan},
                country={P. R. China}}
                
\affiliation[3]{organization={Great Bay Institute for Advanced Study, Mathematics and Information Security Research Center},
                city={Dongguan},
                postcode={523000}, 
                state={Guangdong},
                country={P. R. China}}

\cortext[cor1]{Corresponding author.}

\begin{abstract}
Private Set Intersection (PSI) enables secure computation of set intersections while preserving participant privacy, standard PSI existing protocols remain vulnerable to data integrity attacks allowing malicious participants to extract additional intersection information or mislead other parties. In this paper, we propose the definition of data integrity in PSI and construct two authenticated PSI schemes by integrating Merkle Trees with state-of-the-art two-party volePSI and multi-party mPSI protocols. The resulting two-party authenticated PSI achieves communication complexity $\mathcal{O}(n \lambda+n \log n)$, aligning with the best-known unauthenticated PSI schemes, while the multi-party construction is $\mathcal{O}(n \kappa+n \log n)$ which introduces additional overhead due to Merkle tree inclusion proofs. Due to the incorporation of integrity verification, our authenticated schemes incur higher costs compared to state-of-the-art unauthenticated schemes. We also provide efficient implementations of our protocols and discuss potential improvements, including alternative authentication blocks.
\end{abstract}
\maketitle

\section{Introduction}
Private set intersection (PSI) \cite{Mea86, FNP04} is a cryptographic protocol that enables two parties—often characterized as a client and a server—to compute the intersection of their respective sets, $X$ and $Y$, without revealing any additional information about elements outside of the intersection. Concretely, if one or both parties learn $X \cap Y$, they learn nothing else about their counterpart’s dataset. As one of the most extensively studied protocols in secure multi-party computation (MPC), PSI has emerged as a critical enabler for privacy-preserving applications ranging from genomic data analysis \cite{SCW+18} to private Ad conversion measurement \cite{IKN+17}. 

Even when a PSI protocol provides security against malicious adversaries, it may still permit small-domain probing via probe elements—synthetic items the adversary does not truly possess but inserts solely to test membership. By padding its submitted set with such probe elements, a dishonest participant can mislead the counterparty or infer which items the other party holds, thereby undermining PSI’s intended privacy guarantees.

\subsection{Related Work}
The earliest PSI protocol was built upon the Diffie-Hellman key exchange \cite{Mea86}. Since PSI was formally introduced in \cite{FNP04}, extensive research has been devoted to enhancing both its efficiency and security. Depending on the sizes of the participants’ datasets, PSI protocols can be classified as balanced PSI or unbalanced PSI. Moreover, numerous variants have emerged, including PSI cardinality and threshold PSI. Regardless of the specific functionality sought in a PSI construction, it invariably relies on several fundamental building blocks, including homomorphic encryption, oblivious transfer (OT) and OT-extension, and oblivious pseudo-random functions (OPRF)\cite{MAL23}. 

Following the research line of homomorphic encryption, this method originates from traditional Oblivious Polynomial Evaluation (OPE) protocols \cite{NP06}. By leveraging techniques such as Single Instruction Multiple Data (SIMD) \cite{SV14}, the data sets are encoded into polynomial form, whose coefficients are subsequently encrypted using homomorphic encryption \cite{zheng2}. Participants can then repeatedly invoke membership tests on these polynomials in a homomorphic manner, ultimately deriving the intersection information. However, purely OPE-based protocols often incur higher costs compared to those that rely on alternative cryptographic primitives \cite{KK17}. Some protocols also employ Bloom filters or Cuckoo filters to represent sets more efficiently. By leveraging BFV homomorphic encryption to construct polynomials, along with hash-to-bins techniques and Cuckoo Hashing, \cite{CLR17} offers a cutting-edge PSI protocol in this line of research. 

As for OT- and OPRF-based PSI schemes, although OT can be employed directly for PSI construction, it is often combined with hashing strategies (e.g., simple hashing, permutation-based hashing, or Cuckoo Hashing) for improved efficiency \cite{PSZ14, PSWW18, PSZ18}. In essence, OPRF-based PSI seeks to construct an OPRF by leveraging OT or OT-extension in conjunction with various cryptographic primitives (e.g., hash functions and encryption algorithms) and efficient data structures (e.g., Bloom filters and hash tables). This enables membership tests between sets, ultimately determining the intersection.

A recent line of research is OKVS-based PSI, initially introduced in \cite{PRTY20}. By incorporating a novel linear solver called PaXoS-a binary OKVS built upon an encrypted cuckoo filter and combining it with OT, this approach achieves malicious security in PSI protocols and exhibits robust performance. Building on this, \cite{RS21} proposed an OPRF-based PSI scheme where the OPRF is constructed via Vector Oblivious Linear Evaluation (VOLE). Subsequently, \cite{RR22} presented a more efficient OKVS solution that replaces PaXoS from \cite{PRTY20} and, compared with \cite{RS21}, further adopts subfield-VOLE to enhance performance. These developments represent the current state of the art in PSI.

\subsection{Contribution}
We address integrity in PSI by proposing authenticated PSI, a framework that validates and authenticates each party’s input set to prevent element fabrication and over-claiming attacks. We further define a formal notion of data integrity for PSI, outline new deployment scenarios, and provide two concrete constructions—covering both two-party and multi-party settings—along with reference implementations.
\begin{itemize}
    \item \textbf{Authenticated-Input Integrity for PSI.} We introduce a new malicious-participant scenario for PSI, where parties may alter, inject, or substitute records to bias the intersection or extract additional membership information; in response, we formalize a new integrity notion for PSI that enforces input authenticity and gates access to outputs—each participant is cryptographically bound to a well-formed dataset and verified as untampered throughout the protocol; only parties that pass verification obtain any intersection output, thereby defining Authenticated PSI and preventing data-integrity attacks while strengthening overall protocol security. Besides, we conclude a unifying paradigm for PSI: regardless of the underlying cryptographic primitives, PSI protocols can be cast as instances of a single canonical construction.

    \item \textbf{New constructions.} We proposed several tools to achieve PSI integrity and utilized the representative Merkle Tree to integrate with state-of-the-art two-party PSI (volePSI) and multi-party PSI (mPSI) protocols. Based on these integrations, we developed corresponding authenticated PSI schemes and validated their security and integrity. Our two-party authenticated PSI has communication cost $\mathcal{O}(n \lambda+n \log n)$ which matches the cost of the best unauthenticated schemes \cite{RR22}. As to the multi-party construction, it has communication cost $\mathcal{O}(n \kappa+n \log n)$ which introduces additional complexity arising from the merkle tree inclusion proofs compared to the best multi-party psi \cite{NTY21}.

    \item \textbf{Implementations and Evaluation.}
We realize our authenticated PSI in both two-party and multi-party settings. To highlight advantages beyond throughput, we introduce a targeted threat model performing small-domain probing (including fixed-cardinality substitution) and simulate this integrity attack.

\end{itemize}

\subsection{Notation}
We denote by $\kappa$ the computational security parameter and by $\lambda$ the statistical security parameter. The notation $[a,b]$ represents the set $\{a, a+1, \dots, b\}$, while $[a]$ is used as a shorthand for the set $\{1, \dots, a\}$. The symbol $\langle A, B \rangle$ denotes the inner product between vectors $A$ and $B$. For a set $S$, the notation $s \leftarrow S$ indicates that $s$ is selected as a uniformly random element from the set $S$. Moreover, the symbol $\leftarrow$ is also used to represent the output generated by a specific algorithm defined in this work.
\subsection{Outline}
In \hyperref[sec2]{Section 2}, we will introduce the concept of data integrity, which is essential for safeguarding data security and reliability, and it forms the core motivation for this work. This section will also introduce the Data Manipulation Attack on the  PSI protocols, which compromises data integrity. Such an attack can obscure or mislead the actual intersection data, allowing the malicious participant to gain access to more information than they are entitled to. Besides, merkle tree \cite{Mer89} will be introduced as our core building block which is used to against the attack.

In \hyperref[sec3]{Section 3}, we will provide a unified definition of PSI, along with the associated security and correctness definitions. Additionally, we will introduce a new definition of data integrity for PSI and provide a complete definition of authenticated PSI that is resistant to data manipulation. Furthermore, this section will present the basic building blocks required for our two-party and multi-party PSI protocols, including VOLE \cite{BCGI19}, OKVS \cite{PRTY20, GPR21}, and the key block for ensuring data integrity, the Merkle tree \cite{Mer89}, along with their relevant definitions.

In \hyperref[sec4]{Section 4}, we will present the constructions of our merkle tree enhanced PSI for data integrity for both two-party and multi-party settings, and prove their correctness, data integrity, and security. Additionally, in \hyperref[sec5]{Section 5}, we will provide the implementation and performance analysis of the authenticated PSI, combining volePSI \cite{RS21, RR22} and mPSI \cite{NTY21}, and evaluate it under our small-domain probing threat model to demonstrate the advantages of authenticated PSI. Finally, in \hyperref[sec6]{Section 6}, we will provide a summary of the entire work and discuss alternative approaches for constructing authenticated PSI.

\section{Motivation and Merkle Tree}\label{sec2}
\subsection{Data Integrity}
Standard PSI protocols focus on preserving the privacy of the participants’ sets but provide no guarantees about the authenticity or integrity of the data. If a malicious party (e.g., a server or client) deviates from the protocol, they could manipulate the input sets or the intersection results, leading to incorrect or even harmful outcomes. This lack of integrity protection undermines trust in outputs of PSI and poses critical risks in sensitive applications:
\begin{itemize}
    \item \textit{Medical Research Collaboration:}
Multiple institutions might use PSI to identify shared patient cohorts for joint studies. If an adversary injects falsified patient records into the intersection, researchers could draw erroneous conclusions, jeopardizing treatment efficacy or clinical trial validity.
    \item \textit{Financial Fraud Detection:}
Banks might privately compare transaction blacklists to flag suspicious accounts. A malicious participant could remove entries from the intersection to shield fraudsters or insert innocent users into the result, causing unwarranted account freezes.
    \item \textit{Supply Chain Security:}
Suppliers and manufacturers could use PSI to validate shared components’ origins. Tampering with the intersection might conceal unauthorized sources, enabling counterfeit parts to infiltrate the supply chain and compromise product safety.
\end{itemize}
In non-private applications, integrity can be enforced through established mechanisms. For instance, blockchain systems like Bitcoin use cryptographic hashing and decentralized consensus to ensure the immutability of transaction records, while document management platforms (e.g., DocuSign) employ digital signatures to bind authenticity to verified identities. Similarly, TLS/SSL protocols leverage digital certificates to authenticate website identities and guarantee data integrity in web communications \cite{TLS1.3}. The challenge, however, is to extend these integrity guarantees to private variants of such systems.

\subsection{Data Integrity Attacks on PSI}
Data integrity attacks, often termed Data Manipulation, Data Poisoning, or Data Tampering, constitute a severe yet frequently underestimated threat to numerous organizations. They encompass unauthorized alterations to existing data or the unapproved insertion of fraudulent entries, thereby undermining the accuracy, trustworthiness, and overall security of critical systems.
\begin{figure}[pos=h]
	\centering
	\includegraphics[width=.9\columnwidth]{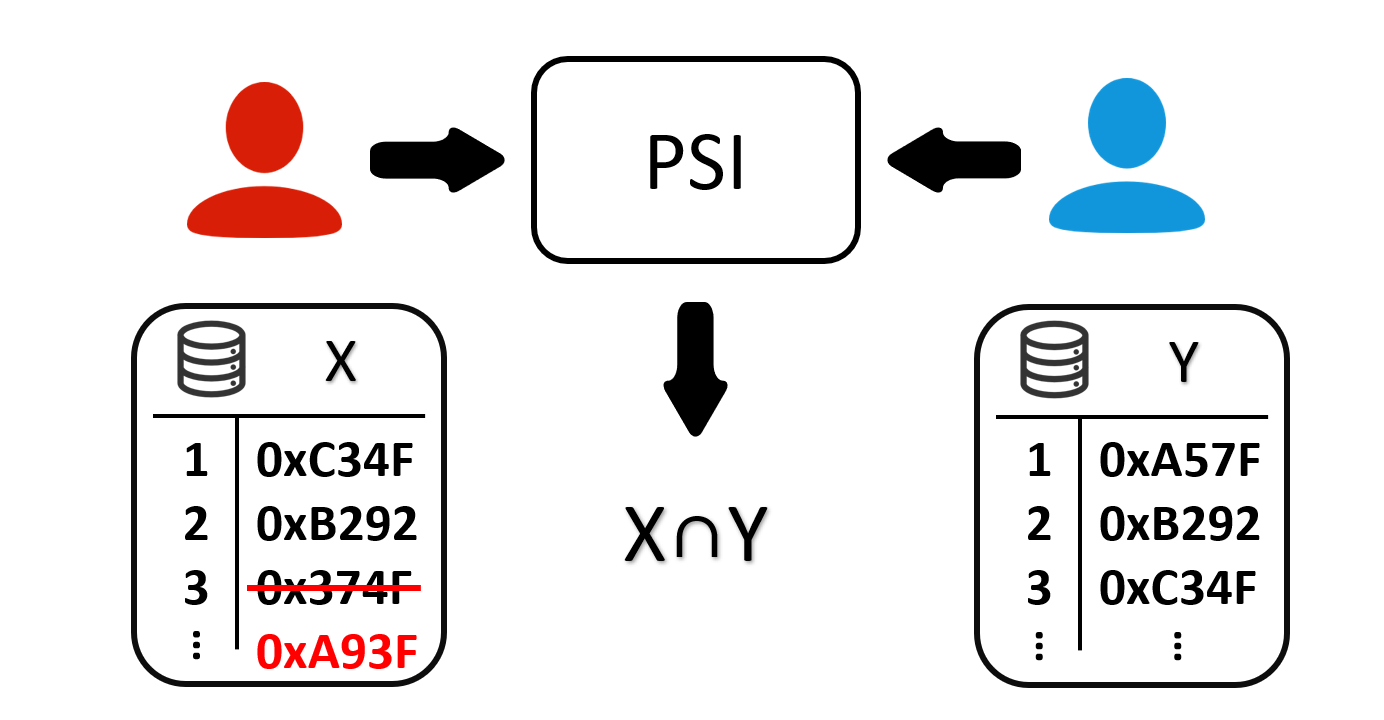}
	\caption{Diagram Illustrating Data Integrity Attacks on PSI}
	\label{fig1}
\end{figure}
Data integrity attacks in PSI occur when a malicious participant deliberately submits forged or manipulated data to obtain unauthorized insight into the intersection or to mislead other participants. The malicious participants can selectively corrupt inputs or infer hidden information through targeted data manipulation. Such attacks can compromise both the correctness of the computed intersection and the privacy of input sets. As shown in figure~\ref{fig1}. For instance, in a scenario where multiple suppliers rely on PSI to determine overlapping inventory for collaborative logistics, a malicious supplier could insert fictitious product identifiers into its dataset. By observing whether these spoofed entries appear in the intersection, the attacker can deduce additional information about the inventory holdings of other supplier, far beyond what should be revealed. This behavior not only subverts the intended privacy guarantees but also undermines the reliability and trustworthiness of the PSI protocol.

\subsection{PSI with Merkle Tree}
Data integrity can be achieved through a variety of cryptographic mechanisms, such as digital signatures, hash chains, or Learning with Errors (LWE)-based schemes. In this work, we adopt Merkle trees \cite{Mer89} as our primary building block for integrity verification, drawing inspiration from prior research \cite{KO97, CN23} that utilizes Merkle trees for authentication in Private Information Retrieval (PIR). 

The evolution of Merkle Trees can be traced back to Ralph Merkle’s 1979 doctoral dissertation, which laid the groundwork for subsequent hash-based authentication mechanisms. However, the formal definition of the Merkle Tree did not appear in published form until 1989 \cite{Mer89}. During the intervening period, Merkle introduced a digital signature scheme \cite{Mer87} built on conventional cryptographic functions, demonstrating for the first time how a hierarchical hashing structure could be applied to verify data integrity. In the 21st century, Michael Szydlo’s work at Eurocrypt 2004 \cite{Szy04} significantly advanced the state of Merkle Trees by reducing the memory requirements of traversal algorithms from $\log^2 n$ to $\log n$, thus greatly enhancing their applicability to large-scale datasets. These landmark contributions collectively propelled Merkle Trees from a theoretical tool into a core cryptographic component.
\begin{figure}[pos=h]
    \centering
    \includegraphics[scale=0.4]{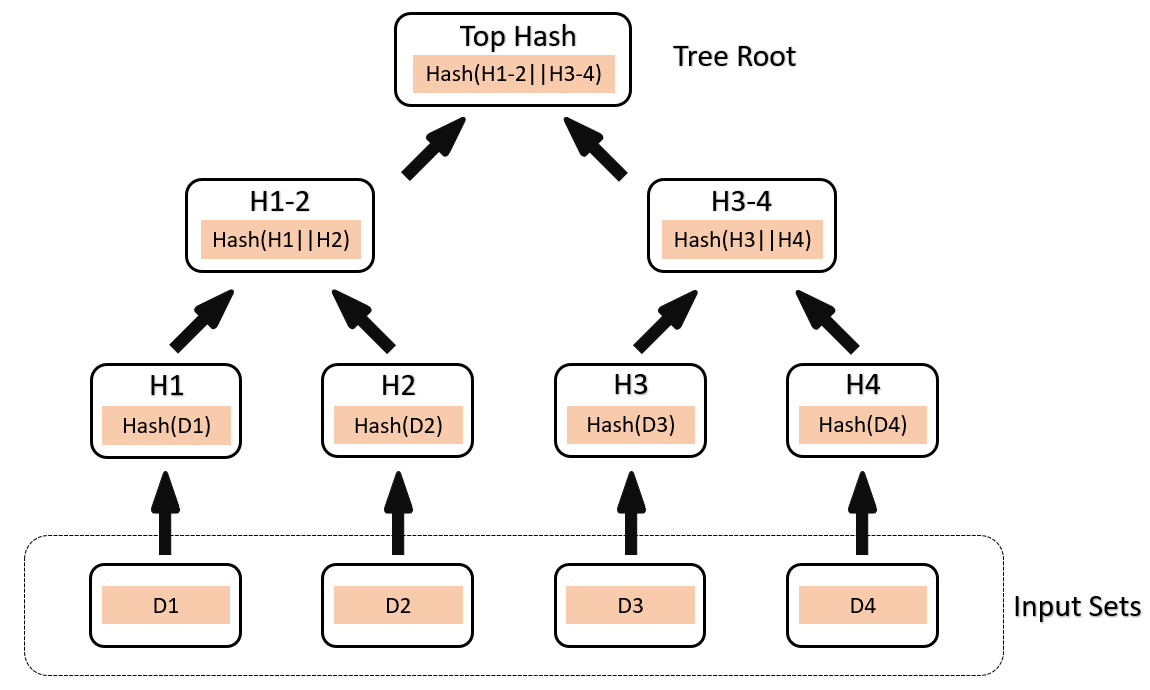}
    \caption{Merkle Tree}
    \label{fig2}
\end{figure}
As shown in figure \ref{fig2}. A Merkle tree ensures data integrity by generating a hash tree where each leaf node represents the hash of an individual data element. The internal nodes store hashes of their child nodes, and the root hash, known as the \textit{Tree Root}, serves as a compact representation of the entire dataset. The inclusion of a data element is verified by traversing the tree from the leaf to the root, ensuring that the data has not been tampered with. Given a set of data blocks \(D_1, D_2, D_3, D_4\), to verify the inclusion of a specific data block, say \(D_3\), a participant would provide the path consisting of the sibling hashes at each level, starting from \(H_3\) and moving upwards to the root which is a tuple $(H_3, H_4, H_{1-2})$. The verification process checks that the computed root hash matches the provided root hash like $root = Hash(H_{1-2}||Hash(H_3||H_4))$. This structure ensures that any modification to a data block would result in a change to its corresponding hash and propagate up the tree, altering the root hash and indicating tampering.

Originally conceived for digital signatures, Merkle trees have evolved into a versatile and efficient mechanism for data integrity verification, assuming a critical role in numerous applications. By hashing data blocks and arranging them into a tree structure, Merkle trees provide concise proofs of authenticity and tamper-resistance. For example, in blockchain systems such as Bitcoin \cite{Nak08} and Ethereum \cite{But13}, they enable rapid verification of transaction histories. In version control systems like Git, a similar hash-based structure is used to track file revisions and ensure data authenticity in distributed environments. Likewise, distributed storage and file-sharing platforms (e.g., IPFS \cite{Ben14}) rely on Merkle trees to guard against data corruption or forgery. Consequently, Merkle trees have become an essential cryptographic primitive for safeguarding data integrity, finding successful applications in blockchain, version control, distributed storage, and beyond.

\section{Preliminaries}\label{sec3}
\subsection{Private Set Intersection (PSI)}
Before delving into the detailed construction of our PSI scheme, it is necessary to provide definitions of correctness and security for both traditional PSI schemes and the authenticated PSI proposed in this paper. These definitions will facilitate the understanding of subsequent work, including the construction of the authenticated PSI and its security proof.

The construction of PSI protocols, regardless of whether they are based on OT and OT-extension, Homomorphic Encryption, or OPRF, follows a similar fundamental logic. Each participant needs to $\mathbf{Transform}$ their input set to a specific format. These transformations often include, but are not limited to, the use of hash functions or the conversion of data into specialized structures, such as OKVS or Hash Graphs. Following these transformations, the participants engage in one or multiple rounds of $\mathbf{Interaction}$ based on the design and structure of the protocol. Finally, using the results from the interaction phase, participants $\mathbf{Reconstruct}$ the intersection of their input.

To avoid redundant descriptions and without loss of generality, we use a unified definition for both two-party PSI and multi-party PSI. When the parties number $n=2$, it corresponds to the definition of two-party PSI.
\begin{definition}\label{PSI}
   For the protocol participants $P_1, \dots, P_n$ where $n \geq 2$, each participant holds an input set $X^1, \dots, X^n$ where $X^i = \{x^{i}_{1},...,x^{i}_{n_{i}}\}$ and $x^{i}_{j} \in \{0, 1\}^{*}$. To compute the intersection $\bigcap_{i=1}^{n} X^i$ of all input sets, a PSI scheme can be represented by three phase: $\mathrm{PSI} = (\mathrm{Transform}, \mathrm{Interaction}, \mathrm{Reco}\\ \mathrm{nstruct})$:
   \begin{itemize}
    \item $\mathbf{Transform}$ \((1^{\lambda}, X) \to (\mathrm{st}, \mathcal{T})\):  Given a security parameter \(\lambda\) and all the input sets $\{X^i\}_{i\in [n]}$ provided by parties $\{P_i\}_{i\in [n]}$, this algorithm outputs a transformation $\mathcal{T}$ of their inputs according to the construction of specific scheme. The algorithm generates a state $\mathrm{st}$, which encapsulates intermediate data (Protocol phase information, randomness) required for subsequent computations .

    \item $\mathbf{Interaction}$ \((1^{\lambda}, \mathrm{st}, \mathcal{T}) \to \mathcal{R}\): This phase takes as input a security parameter \(\lambda\), $\mathcal{T}$ and state information $\mathrm{st}$, after one or multiple rounds of interaction as defined by the protocol, the output $\mathcal{R}$ is obtained. 

    \item $\mathbf{Reconstruct}$ \((\mathrm{st},  \mathcal{R}) \to \mathcal{I}\):  Given the state information $\mathrm{st}$, according to the result of $\mathrm{Interaction}$ phase: $\mathcal{R}$, the parties can obtain the intersection $\mathcal{I}$.
\end{itemize}
\end{definition}
A PSI scheme follows the properties Correctness and Security as follows:
\begin{definition}\label{PSI_Correctness}
$\mathbf{Correctness:}$ For a scheme $\mathrm{PSI} = (\mathrm{Transf}\\ \mathrm{orm}, \mathrm{Interaction}, \mathrm{Reconstruct})$ as defined in $\mathrm{Defintion~\ref{PSI}}$. Given the security parameter $\lambda$ and input set $\{X^i\}_{i \in [n]}$ from $\{P_i\}_{i \in [n]}$ where $n \geq 2$, we say the protocol satisfies the $\mathrm{correctness}$ if the followings holds:
    \[
    \Pr \left[
    \begin{array}{rl}
    & (\mathrm{st}, \mathcal{T}) \leftarrow \mathrm{Transform}(1^{\lambda}, X) \\
    \mathcal{I} = \bigcap_{i=1}^{n} X^i : & \mathcal{R} \leftarrow \mathrm{Interaction}(1^{\lambda}, \mathrm{st}, \mathcal{T}) \\
    & \mathcal{I} \leftarrow \mathrm{Reconstruct}(\mathrm{st}, \mathcal{R}) \\
    \end{array}
    \right] = 1 - \epsilon
    \]
\end{definition}
The correct execution of the protocol ensures that the output $\mathcal{I}$ is equal to the intersection $\bigcap_{i=1}^{n} X^i$ of the participants' input sets. Besides, the soundness error $\epsilon$  depends on the specific PSI scheme. This $\epsilon = negl(\lambda)$ accounts for potential errors arising from probabilistic elements such as hashing, encoding techniques, or cryptographic assumptions used in the PSI protocol. Consequently, ensuring a negligible $\epsilon$ is crucial for achieving high accuracy in real-world applications.

\begin{definition}\label{PSI_Security}
$\mathbf{Security:}$
    Let n parties scheme $\mathrm{PSI} = (\mathrm{Trans}\\ \mathrm{form},\mathrm{Interaction}, \mathrm{Reconstruct})$ as defined in $\mathrm{Defintion~\ref{PSI}}$. For the input sets  $\{X^i\}_{i \in [n]}$, the scheme ensures $\mathrm{security}$ if the following holds:
    \[
    \mathrm{View} = \left\{
    \begin{array}{c l}
    \mathcal{I} : 
    \begin{array}{l}
    (\mathrm{st}, \mathcal{T}) \leftarrow \mathrm{Transform}(1^{\lambda}, X) \\
    \mathcal{R} \leftarrow \mathrm{Interaction}(1^{\lambda}, \mathrm{st}, \mathcal{T})\\
    \mathcal{I} \leftarrow \mathrm{Reconstruct}(\mathrm{st}, \mathcal{R}) 
    \end{array}
    \end{array}
    \right\} 
    \]
For a simulator $\mathcal{S}_{\mathrm{PSI}}$,
    \[
    \mathrm{Sim} = \left\{ 
    \begin{array}{l}
    \mathcal{I}^{\prime} = \{a_i\}_{i \in [n]} \leftarrow \mathcal{S}_{\mathrm{PSI}}(1^{\lambda}, \{0,1\}^{*})
    \end{array}
    \right\} 
    \]
    \[
    \mathrm{View}  \: {\approx}_{c} \: \mathrm{Sim}.
    \]
\end{definition}
The definition essentially means that, from the perspective of an adversary, it is impossible to computationally distinguish between the intersection result obtained through the protocol ($\mathrm{View}$) and a set sampled randomly from a simulator ($\mathrm{Sim}$).

\subsection{Authenticated PSI}
In \cite{DT09}, the notion of Authorized PSI is introduced, bearing some resemblance to authenticated PSI in that it requires verifying that each element in the client’s dataset has been authorized (signed) by a recognized and mutually trusted authority. However, our authenticated PSI differs in its requirement to validate the input datasets of all participants, rather than focusing on one party alone.

Moreover, Authorized PSI places greater emphasis on the authorization of data, allowing for tiered access control to restrict different participants’ data access rights—a feature often employed in international collaborations or proxy-based cloud PSI. This approach strives to balance privacy protection with the need to share data in compliance with organizational or regulatory mandates. In contrast, our authenticated PSI aims specifically to guard against malicious data tampering—ensuring that no participant can undermine the protocol by submitting falsified inputs or glean additional intersection information through fraudulent data manipulation.

Compared to the conventional PSI, authenticated PSI emphasizes the integrity of input set, this subsection presents the definitions for authenticated PSI, including correctness, integrity, and security. Notably, the correctness property aligns with Definition~\ref{PSI_Correctness}. For n parties scheme $\mathrm{PSI} = (\mathrm{Transform}, \mathrm{Interaction}, \mathrm{Reconstruct})$. The parties $\{P_{i}\}_{i \in [n]}$ hold input sets $\{X^i\}_{i \in [n]}$ where $X^i = \{x^{i}_{1},...,x^{i}_{n_{i}}\}$, $x^{i}_{j} \in \{0, 1\}^{*}$ the integrity is defined as follows:
\begin{definition}\label{AuthenticatedPSI_integrity}
    $\mathbf{Integrity:}$ The scheme $\mathrm{PSI} = (\mathrm{Transform}, \\ \mathrm{Interaction}, \mathrm{Reconstruct)}$ with original announced input sets $\{X^i\}_{i \in [n]}$. There is an adversarially modified dataset $X^{i^{*}} \neq X^i$. It satisfies the $\mathrm{integrity}$ as follows:
    \[
    \Pr \left[
    \begin{array}{rl}
    & (\mathrm{st}_{\mathcal{A}}, \mathcal{T}) \leftarrow \mathcal{A}(1^{\lambda}, X^*) \\
    \mathcal{I} \neq \perp : & \mathcal{R} \leftarrow \mathrm{Interaction}(1^{\lambda}, \mathrm{st}_{\mathcal{A}}, \mathcal{T}) \\
    & \mathcal{I} \leftarrow \mathrm{Reconstruct}(\mathrm{st}_{\mathcal{A}}, \mathcal{R}) \\
    \end{array}
    \right] \leq \mathrm{negl(\lambda)}
    \]
\end{definition}
This definition implies that if any participant attempts to manipulate their data to gain additional information, the probability that the process does not terminate is negligible.

\begin{definition}\label{AuthenticatedPSI_security}
     $\mathbf{Security:}$ Based on the $\mathrm{integrity}$ defined in $\mathrm{Definition~\ref{AuthenticatedPSI_integrity}}$, for adversary $\mathcal{A} = (\mathcal{A}_0, \mathcal{A}_1)$, the $\mathrm{Authenticated\,PSI}$ satisfies the $\mathrm{security}$ as follows:
     \[
     \mathrm{REAL}_{\mathcal{A},X,\lambda} = \left\{
     \begin{array}{c l}
     \hat{\mathcal{V}} : & 
     \begin{array}{l}
     (\mathrm{st}_{\mathcal{A}}, \mathcal{T}) \leftarrow \mathcal{A}_0(1^{\lambda}, X^*)\\
     \mathcal{R} \leftarrow \mathrm{Interaction}(1^{\lambda}, \mathrm{st}_{\mathcal{A}}, \mathcal{T}) \\
     \mathcal{I} \leftarrow \mathrm{Reconstruct}(\mathrm{st}_{\mathcal{A}}, \mathcal{R}) \\
      a \leftarrow \mathbbm{1}\{\mathcal{I} \neq \perp \}\\
      \hat{\mathcal{V}} \leftarrow \mathcal{A}_1(\mathrm{st}_{\mathcal{A}}, a)
     \end{array}
     \end{array}
     \right\}
     \]
Similarly, for simulator $\mathcal{S} = (\mathcal{S}_0, \mathcal{S}_1)$:
     \[
     \mathrm{IDEAL}_{\mathcal{A}, \mathcal{S}, X, \lambda} = \left\{
     \begin{array}{c l}
     \mathcal{V} : & 
     \begin{array}{l}
     (\mathrm{st}_{\mathcal{A}}, \mathcal{T}) \leftarrow \mathcal{A}_0(1^{\lambda}, X^*)\\
     (\mathrm{st}_{\mathcal{S}}, \mathcal{I}) \leftarrow \mathcal{S}_0(1^{\lambda}, \mathcal{T}) \\
     a \leftarrow \mathcal{S}_1(\mathrm{st}_{\mathcal{S}}, \mathcal{I}) \\
     \mathcal{V} \leftarrow \mathcal{A}_1(\mathrm{st}_{\mathcal{A}}, a)
     \end{array}
     \end{array}
     \right\}
     \]
    \[
    \mathrm{REAL}_{\mathcal{A},X,\lambda}  \: {\approx}_{c} \: \mathrm{IDEAL}_{\mathcal{A}, \mathcal{S}, X, \lambda}.
    \]
\end{definition}
This definition guarantees that authenticated PSI protocol is computationally indistinguishable between real-world execution and ideal-world simulation, ensuring that no adversary can gain additional information beyond the defined outputs.
\subsection{Merkle Tree}
As the essential verification block, the Merkle Tree is critical for implementing our authenticated PSI, In the context of our work, the Merkle Tree serves as a critical block in combining with PSI to ensure that participants cannot maliciously tamper with their datasets to gain additional information beyond the intended intersection. By leveraging the Merkle Tree's structure, each party commits to their dataset by creating a Merkle Root, which summarizes their data. During the PSI process, each participant must provide proofs (Merkle paths) for their data elements, allowing the other party to verify the integrity of the dataset without directly accessing it. This integration prevents dishonest participants from modifying their inputs in an attempt to infer non-intersecting elements, thereby enhancing the security and trustworthiness of the authenticated PSI protocol. We now begin by introducing the algorithms and processes involved in the Merkle Tree.    

\begin{definition}\label{MerkleTree}      
We represent the Merkle tree as a scheme consisting of three algorithms: $\mathrm{Merkle} = (\mathrm{Root}, \mathrm{GenPath}, \mathrm{Verify})$, parameterized by  \(\ell_{\text{root}}, \ell_{\pi} \in \mathbb{N}\) which represent the root length and tree-path length respectively. For an input set $X=\{x_{1},...,x_{n} \}, x_{i}\in \{0,1\}^*$. Given a security parameter \(\lambda\), the scheme is composed of two probabilistic algorithms and one deterministic algorithm:
\begin{itemize}
    \item $\mathbf{Root}$ \((1^{\lambda}, X) \to root\):  Given a security parameter \(\lambda\) and dataset $X=\{x_{1},...,x_{n} \}$, this probabilistic algorithm outputs a tree \(root \in \{0,1\}^{\ell_{\text{root}}}\) representing the set.
    \item $\mathbf{GenPath}$ \((1^{\lambda}, X, x_i) \to \pi_i\): The deterministic algorithm takes as input a security parameter \(\lambda\), a input set \(X\), and a element \(x_i \in \{0,1\}^{*}\), outputs a unique tree path \(\pi_i \in \{0,1\}^{\ell_{\pi}}\) which is used to verify the inclusion.
    \item $\mathbf{Verify}$ \((root, i, \pi_i) \to \{ \perp,1\}\): Given a \(root \in \{0,1\}^{\ell_{\text{root}}}\) and a tree path \(\pi_i \in \{0,1\}^{\ell_{\pi}}\) for $x_i$, this algorithm outputs \(1\) if \(\pi_i\) proves that the corresponding element \(x_i\) is included in the merkle tree represented by \(root\); otherwise, verification of the path failed, the program terminates.
\end{itemize}
\end{definition}
  
The main definitions of the Merkle Tree are categorized into three key properties: Correctness, Uniqueness, and Soundness as follows.

\begin{definition}\label{MerkleTree_Correctness}
$\mathbf{Correctness:}$ For a Merkle tree scheme defined in $\mathrm{Definition~\ref{MerkleTree}}$, $\mathrm{Merkle = (\mathrm{Root}, \mathrm{GenPath}, \mathrm{Verify})}$ is parameterized by a root length \(\ell_{\text{root}} \in \mathbb{N}\) and a corresponding tree path length \(\ell_{\pi} \in \mathbb{N}\). For an input set $X=\{x_{1},...,x_{n} \}$ with $x_{i}\in \{0,1\}^*$, we say that the $\mathrm{Merkle}$ scheme satisfies $\mathrm{correctness}$ if, for all \(i \in [n]\), the following holds:
\hspace{-0.5cm}
\[
\Pr \left[
\begin{array}{rl}
& root \leftarrow \mathrm{Root}(1^{\lambda},X) \\
b = 1 : & \pi_i \leftarrow \mathrm{GenPath}(1^{\lambda}, X, x_i) \\
& b \leftarrow \mathrm{Verify}(root, i, \pi_i) \\
\end{array}
\right] = 1
\]
\end{definition}
If an element is indeed present in the dataset, the proof generated by the Merkle tree should be successfully verified. The following properties, uniqueness and soundness, emphasize the one-to-one correspondence between $x_{i}$ and its tree path in the merkle tree.

\begin{definition}\label{MerkleTree_Uniqueness}
$\mathbf{Uniqueness:}$ Let $\mathrm{Merkle} = (\mathrm{Root},  \mathrm{GenPath}\\ , \mathrm{Verify})$ be a Merkle-tree scheme as specified in $\mathrm{Definition~\ref{MerkleTree}}$, with parameters for the root \(\ell_{\text{root}} \in \mathbb{N}\) and the tree path length \(\ell_{\pi} \in \mathbb{N}\). Given an input set $X=\{x_{1},...,x_{n} \}$ with $x_{i}\in \{0,1\}^*$, let $\mathcal{A}$ be an efficient adversary. $\mathrm{Merkle}$ ensures $\mathrm{uniqueness}$ if the following holds:
\[ 
\Pr \left[
\begin{array}{rl}
& (X, x_i, \pi_i, \pi_i') \leftarrow \mathcal{A}(1^{\lambda}, n) \\
&  \pi_i \neq \pi_i': \\
b = b' = 1 : & root \leftarrow \mathrm{Root}(1^{\lambda},X) \\
& b \leftarrow \mathrm{Verify}(root, i, \pi_i) \\
& b' \leftarrow \mathrm{Verify}(root, i, \pi_i')
\end{array}
\right] \leq \mathrm{negl(\lambda)}
\]
\end{definition}
This definition ensures that each element \( x_i \) in the input set \( X \) has a unique tree path \(\pi_i\). It prevents an adversary from finding two distinct paths that authenticate the same element under the same root.

\begin{definition}\label{MerkleTree_Soundness}
$\mathbf{Soundness:}$ A scheme $\mathrm{Merkle} = (\mathrm{Root}, \mathrm{Gen}\\ \mathrm{Path}, \mathrm{Verify})$ as defined in $\mathrm{Definition~\ref{MerkleTree}}$, with the root length and tree path length $\ell_{\text{root}}, \ell_{\pi} \in \mathbb{N}$. For the input set $X=\{x_{1},...,x_{n} \}, x_{i}\in \{0,1\}^*,$. let $\mathcal{A}$ be an adversary. We say that $\mathrm{Merkle}$ satisfies $\mathrm{Soundness}$ if the following holds:
\[
\Pr \left[
\begin{array}{c l}
b = 1 : & 
\begin{array}{l}
(X, x_{i}, x_{i}^{*}, \pi_i^{*}) \leftarrow \mathcal{A}(1^{\lambda}, n) \\
x_i \neq x_i^*: \\
root \leftarrow \mathrm{Root}(1^{\lambda},X) \\
b \leftarrow \mathrm{Verify}(root, i, \pi_i^{*})
\end{array}
\end{array}
\right] \leq \mathrm{negl(\lambda)}
\]
\end{definition}
This property ensures that an tree path for any arbitrary element \( x_i^* \) cannot be used to verify a different element \( x_i \). It guarantees that each path corresponds exclusively to its specific element.\\

\subsection{Oblivious Key-Value Store (OKVS)}
The OKVS creates a special data structure which enables secure storage and retrieval of key-value pairs without revealing information about the keys or values, providing a critical building block for privacy-preserving data operations. 

Before defining OKVS, it is essential to first understand the basic concept of a Key-Value Store (KVS), which serves as the foundation for its construction.
\begin{definition}\label{KVS}
A Key-Value Store (KVS) scheme is defined as $\mathrm{KVS} = (\mathrm{Encode_H}, \mathrm{Decode_H)}$, where \(H\) represents a set of Hash functions which indicate a specific mapping, \(\mathcal{K}\) is a set of keys, and \(\mathcal{V}\) is a set of values. The input to the KVS scheme consists of a set of key-value pairs \(\{(k_1,v_1),\dots,(k_n,v_n)\}\) where  \((k_i,v_i) \in K \times V\) and \(K \subset \mathcal{K}\),\(V \subset \mathcal{V}\). A KVS scheme provides the following two algorithms:
\begin{itemize}
     \item $\mathbf{Encode}_H$ $(K,V) \to \{S,\perp\}$:  
    This algorithm encodes a set of key-value pairs into a single object \(S\). If the encoding process fails, it outputs an error indicator \(\perp\), which occurs with statistically small probability.
    
    \item $\mathbf{Decode}_H$ $(S, k) \to w$:  
    Given an encoded object \(S\) and a key \(k \in \mathcal{K}\), this algorithm retrieves and outputs the value \(w = v \in \mathcal{V}\) if $(k, v) \in \{(k_i, v_i)\}_{i\in[n]}$. Otherwise, it outputs a randomly generated value \(w = r\).
\end{itemize}
\end{definition}
The probability of the KVS $\mathrm{Encode}$ algorithm outputting \(\perp\) depends on the hash functions used in the data structure, i.e., the specific mapping relationship. For instance, if the key-value pairs are simply stored in a polynomial structure, the $\mathrm{Encode}$ algorithm will always succeed. According to the Lagrange interpolation theorem, \(n\) key-value pairs can always be represented and reconstructed using a polynomial of degree \(n-1\).

Next, we will present the definition of the oblivious version of KVS and notice that the subscript $H$ in Encode and Decode will be omitted in the following content.
\begin{definition}\label{OKVScorrectness}
$\mathbf{Correctness:}$ For a KVS scheme $\mathrm{KVS} = (\mathrm{Encode,Decode})$, given the key-value pairs set $K \times V$ and security parameter $\lambda$, the scheme satisfies $\mathrm{Correctness}$ if the following holds:
\[
\Pr \left[\mathrm{Encode}(K,V) = \perp \right] \leq negl(\lambda)
\]
and
\[\Pr \left[ 
    (k_i', v_i') = (k_j, v_j) \ \Bigg| \ 
    \begin{aligned}
        & S \leftarrow \mathrm{Encode}(K, V), \\
        & v_i' \leftarrow \mathrm{Decode}(k_i', S) \\
        & \quad\ \text{s.t. } S \neq \perp \ \land \ k_i' = k_j \in K
    \end{aligned}
\right] = 1\]
\end{definition}
The correctness of the scheme encompasses the accuracy of both the $\mathrm{Encode}$ and $\mathrm{Decode}$ algorithms. From the perspective of $\mathrm{Encode}$, the probability that it correctly outputs a valid data structure must be smaller than a negligible function based on the security parameter. In the context of OKVS schemes, this negligible function is often $2^{-\lambda}$. As noted in \cite{RR22} and \cite{GPR21}, this probability can reach $2^{-40}$. For $\mathrm{Decode}$, correctness implies that, given a data structure $S$ generated by a properly executed program, decoding any key from the original key-value pairs $K$ should reliably produce the corresponding value $v$. Next, we formally present the definition of OKVS.
\begin{definition}\label{OKVS}
$\mathbf{Obliviousness:}$ A Key-Value Store (KVS) is considered an Oblivious KVS (OKVS) if, for any distinct sets 
$K^0 = \{k_1^0, \dots, k_n^0\}$ 
and 
$K^1 = \{k_1^1, \dots, k_n^1\}$, 
when the $\mathrm{Encode}$ algorithm does not output $\perp$ for either $K^0$ or $K^1$, the distribution 
$D_{\text{Encode}, K^0}$ 
is computationally indistinguishable from 
$D_{\text{Encode}, K^1}$. 
Here, $D_{\text{Encode}, K}$ denotes the output distribution of the $\mathrm{Encode}$ algorithm for the set $K$ as follows:
\[
D_{\text{Encode}, K} = \left\{
\begin{array}{c l}
\mathrm{Encode}(K,V) : 
\begin{array}{l}
v_i \leftarrow V, i \in [n] \\
V = \{v_{1}, \dots,v_{n}\} 
\end{array}
\end{array}
\right\} 
\]
\[
D_{\text{Encode}, K^{0}} \: {\approx}_{c} \: D_{\text{Encode}, K^{1}}.
\]
\end{definition}
$D_{\text{Encode}, K^0}$ and $D_{\text{Encode}, K^1}$ are computationally indistinguishable, meaning that an adversary cannot infer whether the output of the $\mathrm{Encode}$ algorithm originates from $K^0$ or $K^1$. This property is precisely what defines obliviousness and ensures strong privacy guarantees. 

Next, we will specifically define a particular type of OKVS. This is because many existing works are based on this type of OKVS, achieving impressive efficiency results, including the scheme \cite{RR22,NTY21}.

\begin{definition}\label{LinearOKVS}
An OKVS is considered \textbf{linear} over a field $\mathbb{F}$ if the value set $\mathcal{V}$ consists of elements in $\mathbb{F}$, the output of the Encode algorithm is represented as a vector $S \in \mathbb{F}^m$, and the $\mathrm{Decode}$ function is expressed as:
\[
\mathrm{Decode}(S, k) = \langle d(k), S \rangle \overset{\text{def}}{=} \sum_{j=1}^m d(k)_j S_j = v
\]
or
\[
\begin{bmatrix}
d(k_1) \\ 
d(k_2) \\ 
\vdots \\ 
d(k_n)
\end{bmatrix}
S^\top =
\begin{bmatrix}
v_1 \\ 
v_2 \\ 
\vdots \\ 
v_n
\end{bmatrix}
\]

where $d: \mathcal{K} \to \mathbb{F}^m$ is a specific function. Consequently, the  $\mathrm{Decode}$ operation for a fixed $k$ is a linear map from $\mathbb{F}^m$ to $\mathbb{F}$ which is typically defined under the hash function $H$.
\end{definition}
Notice that the Linear OKVS satisfies the obliviousness property defined in Definition~\ref{OKVS}, since for all distinct $\{k_1, \dots, k_n\}$, the set $\{d(k_1), \dots, d(k_n)\}$ is linearly independent with overwhelming probability, and $v_i$ and $S$ are uniformly distributed.

There is a specialized form of a linear OKVS named \textbf{binary OKVS}, which is defined over a field $\mathbb{F}$, where the vectors $d(k)$ are limited to binary values within $\{0, 1\}^m \subseteq \mathbb{F}^m$. In this construction, the Decode operation simplifies to summing specific entries from the vector $S$.

For practical implementations, we often work with the field $\mathbb{F} = GF(2^\ell) \cong \{0, 1\}^\ell$, where addition corresponds to XOR operations on binary strings. This variant, known as a \textbf{probe and XOR of strings (PaXoS)} data structure, was introduced in \cite{PRTY20} and has been widely adopted for its efficiency in handling binary values~\cite{RS21, NTY21}.The prior work \cite{RS21} of scheme \cite{RR22} is based on PaXoS and extends it into a new version called XoPaXoS for its Circuit PSI construction. XoPaXoS offers several improvements over the original PaXoS, including enhanced randomization, and improved practical utility.

\subsection{Vector Oblivious Linear Evaluation (VOLE)}
VOLE is an extension of Oblivious Linear Evaluation (OLE), a secure two-party protocol. While OLE allows a receiver to learn a secret linear combination of two elements held by the sender, VOLE expands this functionality to vectors. VOLE is particularly useful in cryptographic applications as it significantly reduces the cost of generating multiple OLE instances, this efficiency makes VOLE an essential tool in various advanced cryptographic schemes such as zero-knowledge proofs and PSI \cite{BCGI19, SGRR19, WKYW20, YWL+20, RS21, RR22}. 

A common method for generating VOLE correlations involves reducing the problem to random string OT. Random string OT correlations can be efficiently compressed using pseudorandom generators (PRGs). \cite{BCGI19} proposes simple and efficient constructions of VOLE correlation generators. This work leverages Function Secret Sharing (FSS) with Distributed Point Functions (DPF) to construct Function Secret Sharing for Multi-Point Functions (MPFSS), and subsequently bases its security on the Learning Parity with Noise (LPN) assumption over large fields. Unlike the Learning With Errors (LWE) assumption, the LPN assumption restricts the noise to have a low Hamming weight \cite{zheng}. This assumption can also be equivalently described as requiring that the syndrome of a randomly chosen low-weight noise vector appears pseudo-random. 
\begin{definition}\label{VOLE}
A $\mathrm{VOLE}$ scheme is defined by two probabilistic polynomial-time (PPT) algorithms, denoted as $\mathrm{VOLE} = (\mathrm{GenSeed}, \mathrm{Extend})$, which are parameterized by the security parameter $\lambda$ and operate over a finite field $\mathbb{F}$ as follows:

\begin{itemize}
    \item $\mathbf{GenSeed}$ $(1^{\lambda}, \mathbb{F}) \to \{seed_{0},seed_{1}\}$:  
    Given a security parameter $\lambda$, and with the field of output pre-specified, the algorithm generates a pair of seeds $(seed_0, seed_1)$ for $\mathrm{Receiver}$ and $\mathrm{Sender}$, where $seed_1$ includes a scalar $x \in \mathbb{F}$.
    
    \item $\mathbf{Extend}$ $(seed_{\sigma}) \to \mathbf{Z}_{\sigma}$:  
    Given a party index $\sigma \in \{0,1\}$ and the corresponding seed $\mathrm{seed}_{\sigma}$. For $\sigma = 0$ $\mathrm{(Receiver)}$, it outputs a pair of random vectors $\mathbf{Z}_0 = (\mathbf{a}, \mathbf{c}) \in \mathbb{F}^m \times \mathbb{F}^m$. For $\sigma = 1$ $\mathrm{(Sender)}$, it outputs a scalar-vector pair $\mathbf{Z}_1 = (\mathbf{b}, x) \in \mathbb{F}^m \times \mathbb{F}$, where $x \in \mathbb{F}$ is a scalar. These outputs satisfy the linear combination $\mathbf{c} = \mathbf{a} \cdot x + \mathbf{b}$.
\end{itemize}
\end{definition}
A VOLE scheme always possesses the properties Correctness and Security as follows.
\begin{definition}\label{VOLE_Correctness}
$\mathbf{Correctness:}$
    For a VOLE scheme $\mathrm{VOLE} = (\mathrm{GenSeed}, \mathrm{Extend})$ as defined in $\mathrm{Definition~\ref{VOLE}}$ which is parametrized by field $\mathbb{F}$ and security parameter $\lambda$, we say $\mathrm{VOLE}$ scheme satisfies correctness if:
    
    \vspace{0.2cm}
    \hspace{-4mm}\resizebox{0.5\textwidth}{!}{$
    \Pr \left[
    \begin{array}{rl}
    & \{seed_{0},seed_{1}\} \leftarrow \mathrm{GenSeed}(1^{\lambda}, \mathbb{F}) \\
    \mathbf{c} = \mathbf{a} \cdot x + \mathbf{b}:& (\mathbf{a}, \mathbf{c}) \leftarrow \mathrm{Extend}(seed_{0}) \\
    & (\mathbf{b}, x) \leftarrow \mathrm{Extend}(seed_{1}) \\
    \end{array}
    \right] = 1 - \epsilon
    $}\vspace{0.2cm}
\end{definition}
This definition signifies that the seeds generated by a well-defined algorithm, along with the corresponding outputs derived from these seeds, are guaranteed to satisfy the linear relationship. As for the security of VOLE, it is generally divided into two parts: the security of the GenSeed algorithm and the security of the Extend algorithm.

\begin{definition}\label{VOLE_Security}
For any polynomial-time adversary $\mathcal{A}$, attempting to extract additional information from the output $seed_0$ of the $\mathrm{GenSeed}$ algorithm, the resulting distribution remains computationally indistinguishable, as described below:
    \[
    D_{\text{GenSeed}} = \left\{
    \mathcal{A}(seed_0): 
    (seed_0, seed_1) \leftarrow \mathrm{GenSeed}(1^{\lambda})   
    \right\} 
    \]
    Since the $\mathrm{GenSeed}$ algorithm in VOLE is a PPT algorithm, its outputs differ with each execution. Specifically, the scalar included in $seed_1$, denoted as x, will not be equal to $x'$ in two separate runs. Consequently, an adversary cannot recognize the distribution of the seeds, nor derive any meaningful information from $seed_0$.
    
    As to the security of $\mathrm{Extend}$ algorithm, for any polynomial-time adversary $\mathcal{A}$, it holds that
    {\small\[
    D_{\text{Extend}} = \left\{
    \begin{array}{c l}
    \mathcal{A}(\mathbf{a}, \mathbf{c}, seed_1): 
    \begin{array}{l}
    (seed_0, seed_1) \leftarrow \mathrm{GenSeed}(1^{\lambda}) \\
    (\mathbf{a}, \mathbf{c}) \leftarrow \mathrm{Extend}(seed_0)
    \end{array}
    \end{array}
    \right\} 
    \]}
    {\small\[
     D_{\text{Extend}}^{\prime} = \left\{
    \begin{array}{c l}
    \mathcal{A}(\mathbf{a}, \mathbf{c}, seed_1): 
    \begin{array}{l}
    (seed_0, seed_1) \leftarrow \mathrm{GenSeed}(1^{\lambda}) \\
    (\mathbf{b}, x) \leftarrow \mathrm{Extend}(seed_{1})\\
    \mathbf{a} \leftarrow \mathbb{F}^{m}, \mathbf{c} \leftarrow \mathbf{a} \cdot x + \mathbf{b}
    \end{array}
    \end{array}
    \right\} 
    \]}
    \[
    D_{\text{Extend}} \: {\approx}_{c} \: D_{\text{Extend}}^{\prime}.
    \]
\end{definition} 
The overall security ensures that, whether the algorithm is GenSeed or Extend, an adversary cannot distinguish between their output distributions within polynomial time and can not extract any additional information.

\section{PSI for Data Integrity}\label{sec4}
\subsection{Two-party PSI}
\begin{tcolorbox}[colback=white]\label{construction1}
\textbf{Construction 1: Authenticated volePSI} \\
For our two parties PSI scheme which is denoted as $\mathrm{PSI} = (\mathrm{Transform}, \mathrm{Interaction}, \mathrm{Reconstruct})$ defined in Definition~\ref{PSI}. Given the statistical security parameter $\lambda$ and computational security parameter $\kappa$. The construction is parametrized by a size of input sets $n_{x}, n_{y} \in \mathbb{N}$ where $X = \{x_1,\dots,x_{n_x}\}$ and $Y = \{y_1,\dots,y_{n_y}\}$ and $x_i, y_i \in \{0, 1\}^{*}$ for Receiver and Sender respectively. The parties need to announced the root of their sets where $root \leftarrow \mathrm{Merkle.Root}(1, X)$. There is a field $\mathbb{B}$ with extension $\mathbb{F}$ such that $\lvert \mathbb{F} \rvert = \mathcal{O}(2^\kappa)$ and $\lvert \mathbb{B} \rvert \ge 2^{\lambda + \log_{2}(n_x)+\log_{2}(n_y)}$. Let $\mathrm{H}^\mathbb{B} : \{0,1\}^* \to \mathbb{B}, \mathrm{H}^\circ : \{0,1\}^* \to \{0,1\}^\text{out}$ be random oracles in the set of Hash functions from OKVS where $\text{out}=\lambda + \log_{2}(n_x n_y)$. The construction is combined with \textbf{Merkle Tree}, \textbf{OKVS} and \textbf{VOLE} as defined in Definition~\ref{MerkleTree}, \ref{OKVS}, \ref{VOLE}. (Notice that the computational security parameter $\kappa$ is used in the parameter selection for component of algorithm such that the randomness of Hash functions, and st is used to record the state information of procedure, for simplicity, these details are omitted) 
\paragraph{\underline{$\mathbf{Transform}$ \((1^{\lambda}, X, Y) \to (\mathrm{st}, \mathcal{T})\)}}
\begin{enumerate}
\item The Receiver and Sender generate tree paths for their input sets:

\vspace{1mm}
\resizebox{\linewidth}{!}{$
\left\{
\begin{aligned}
    \Pi_{X} &= \{\pi_{x_i}\}_{i\in [n_x]}, & \pi_{x_i} &\leftarrow \mathrm{Merkle.GenPath}(1^\lambda, X, x_i), \\
    \Pi_{Y} &= \{\pi_{y_i}\}_{i\in [n_y]}, & \pi_{y_i} &\leftarrow \mathrm{Merkle.GenPath}(1^\lambda, Y, y_i).
\end{aligned}
\right.$}\vspace{1mm}
    \item The Receiver computes: 
    
    \resizebox{\linewidth}{!}{$\mathbf{P} \leftarrow \mathrm{OKVS.Encode}(L), L = \{(x, \mathrm{H}^{\mathbb{B}}(x)) \mid x \in X\}.$}\vspace{0.1cm}
    $\mathcal{T} \leftarrow \mathbf{P}.$
    \item The Receiver and Sender exchange the $(root_R, \Pi_{Y})$ and $(root_S, \Pi_{X})$.
\end{enumerate}

\paragraph{\underline{$\mathbf{Interaction}$ \((1^{\lambda}, \mathrm{st}, \mathcal{T}) \to \mathcal{R}\)}}
\begin{enumerate}
\item The Receiver and Sender process the verify procedure such that:
\[
\left\{
\begin{aligned}
   \mathcal{C}_{X} &\leftarrow \mathrm{Merkle.Verify}(root_X, \Pi_X), \\
   \mathcal{C}_{Y} &\leftarrow \mathrm{Merkle.Verify}(root_Y, \Pi_Y).
\end{aligned}
\right.
\]
If $\mathcal{C}_{X}$ or $ \mathcal{C}_{Y}$ contain $\perp$, output $\mathcal{R} = \perp$ and abort the procedure.
\item The Receiver and Sender invoke:
 \[(seed_{0},seed_{1}) \leftarrow \mathrm{VOLE.GenSeed}(1^{\lambda}, \mathbb{F}).\] with output length $m = \lvert \mathbf{P} \rvert$, the Receiver receive $seed_0$ while Sender receive $seed_1$.
\item The Receiver and Sender computes respectively:
\[
\left\{
\begin{aligned}
    (\mathbf{A}, \mathbf{C}) \leftarrow \mathrm{VOLE.Extend}(seed_0), \\
    (\mathbf{B}, \Delta) \leftarrow \mathrm{VOLE.Extend}(seed_1).
\end{aligned}
\right.
\]
where $\mathbf{A}, \mathbf{B}, \mathbf{C} \in \mathbb{F}^{m}$ and $\Delta \in \mathbb{F}$ and $\mathbf{C} = \mathbf{A} \cdot \Delta + \mathbf{B}$.
  \item The Receiver sends $\mathbf{A}^\prime = \mathbf{A} + \mathbf{P} \in \mathbb{B}^{m}$ to sender and Sender computes $\mathbf{B}^\prime = \mathbf{B} + \mathbf{A}^\prime \cdot \Delta \in \mathbb{F}^{m}$.
  \item The Sender outputs and sends: 
  
  \vspace{0.15cm}
  \hspace{-0.3cm}\resizebox{1.1\linewidth}{!}{
  $\mathcal{R} = \{\mathrm{H}^\circ\big(\mathrm{OKVS.Decode}(\mathbf{B'}, y) - \Delta \cdot \mathrm{H}^\mathbb{B}(y)\big) \;|\; y \in Y\}.$}
\end{enumerate}

\paragraph{\underline{$\mathbf{Reconstruct}$ \((\mathrm{st},  \mathcal{R}) \to \mathcal{I}\)}}
\begin{enumerate}
    \item The Receiver computes:
    \[
    \mathcal{R}^{\prime} = \{\mathrm{H}^\circ\big(\mathrm{OKVS.Decode}(\mathbf{C}, x)\big) \;|\; x \in X\}.
    \]
    \item Output $\mathcal{I} \leftarrow \mathcal{R}\bigcap\mathcal{R}^{\prime}$.
\end{enumerate}
\end{tcolorbox}

\subsection{Multi-party PSI}
Before the description of our construction on authenticated multiparty PSI based on the $t$ collusion scheme in \cite{NTY21}, the unconditional zero sharing protocol proposed in \cite{KMP+17} will first be introduced as a tool in the scheme. As the name suggests, the zero shareing can share $n$ parts with XOR sum equals 0.
\begin{tcolorbox}[colback=white]\label{zerosharing}
\textbf{Zero-Sharing}\\
For $n$ parties in the scheme denoted as $P_1,\dots,P_n$ hold the input sets $\{X^i\}_{i \in [n]}$ where the parties have the same size of input $x$. Let $F_{k}(x)$ be a pseudo-random function (PRF), where $k$ is the secret key associated with the function. 

\begin{enumerate}
\item Each party $P_i$ is required to generate and transmit a random seed $k_{i,j}$ to every subsequent participant $P_j$, where $j \in [i+1, n]$. Consequently, each party $P_i$ obtains a key set $K_i$ consisting of $n-1$ elements, specifically
$
K_i = \{\, k_{1,i},\ \dots,\ k_{i-1,i},\ k_{i,i+1},\ \dots,\ k_{i,n} \,\}.
$
\item Each party $P_i$ computes share $S(K_i,x)$ for every $x$ in their input set $X^{i}$ as follows:
\[
S(K_i, x) = \left( \bigoplus_{j < i} F_{k_{j,i}}(x) \right) \oplus \left( \bigoplus_{j > i} F_{k_{i,j}}(x) \right)
\]
\end{enumerate}
\end{tcolorbox}

Subsequently, if $n$ participants jointly possess a data $x$, the XOR of their shares $S(K_i, x)$ associated with this data equals zero which is $\bigoplus_{i=1}^n S(K_i, x) = 0$. Since as for the shared data $x$, for different participants $P_a$ and $P_b$, the pseudo-random function output $F_{k_{a,b}}(x)$ appears exactly twice: once in $S(K_a, x)$ from $P_a$ and once in $S(K_b, x)$ from $P_b$, such terms always pair up and XOR to $0$. By continuing this reasoning, the final result simplifies to $\bigoplus_{i=1}^n S(K_i, x) = 0$.

We provide a brief introduction to the Oblivious Programmable PRF (OPPRF), which serves as a crucial component in the mPSI scheme \cite{NTY21}. OPPRF is an enhanced variant of the Oblivious Pseudorandom Function (OPRF) functionality. While it retains the core properties of OPRF, OPPRF introduces the capability for the sender \( S \) to initially supply a predefined set of points \( \mathcal{P} = \{(x_1, y_1), \dots, (x_m, y_m)\} \) that are programmed into the pseudorandom function. The receiver \( R \) obtains the secret output based on the input as follows:\\
\[
\mathrm{OPPRF}_{S}(x) =
\begin{cases}
y_i & \quad \text{if } x \in \{x_1, \dots, x_m\} \\
\$ & \quad \text{otherwise}
\end{cases}
\]

This expression demonstrates that the receiver will obtain the corresponding pre-programmed value \( y_i \) if the input \( x \) is included in \( \mathcal{P} \). Otherwise, a random value \( \$ \) will be generated. By enabling the sender to define specific input-output pairs, OPPRF enhances the flexibility and functionality of traditional OPRF in privacy-preserving computations.
\begin{tcolorbox}[colback=white]\label{construction2}
\textbf{Construction 2: Authenticated mPSI}\\ 
Our multi parties PSI scheme follows the same notation as \hyperref[construction1]{Construction 1} from Definition~\ref{PSI}. Given the statistical security parameter $\lambda$ and computational security parameter $\kappa$. For parties $P_1,\dots,P_n$ with input sets $X^{1},\dots,X^{n}$, the construction is parametrized by a size of input sets $n_{l} \in \mathbb{N}$ where $X^{i} = \{x^{i}_1,\dots,x^{i}_{n_l}\}, i\in [n]$ for each party $P^{i}$. The root of their sets where $root_{i} \leftarrow \mathrm{Merkle.Root}(1, X^i)$. The construction is combined with \textbf{Merkle Tree}, \textbf{OKVS} as defined in Definition~\ref{MerkleTree}, \ref{OKVS} where specially the OKVS construction is \textbf{Paxos} from \cite{PRTY20}. The proposed scheme is an n-party PSI protocol that can tolerate up to $t$ collusion.
\paragraph{\underline{$\mathbf{Transform}$ \((1^{\lambda}, X, Y) \to (\mathrm{st}, \mathcal{T})\)}}
\begin{enumerate}
    \item All the parties $P_1,\dots,P_n$ generate tree paths for their input sets:\\
    \vspace{0.1cm}
    \hspace{-0.3cm}\resizebox{1.1\linewidth}{!}{
    $\Pi_{X^i} = \{\pi_{x^{i}_j}\}_{j\in [n_l]},  \pi_{x^{i}_j} \leftarrow \mathrm{Merkle.GenPath}(1^{\lambda}, X^i, x^{i}_j)$}
    \vspace{-5mm}
    \item Let $v = n - t$ and divide all \( n \) participants into two groups and a central coordinator: group A consists of \( \{P_1, \dots, P_{v-1}\} \), group B consisting of \( \{P_{v+1}, \dots, P_n\} \), and the central coordinator \( P_v \).
    \item Party $P_i$ in group A generates and sends the keys $\{k_{i}^j\}$ for the party $P_j$ in group B.
    \item Party $P_i$ in group A sends the $T_i$ to center $P_v$ where:\\
    \vspace{0.1cm}
    \hspace{-0.4cm}\resizebox{1.1\linewidth}{!}{
    $T_i \leftarrow \mathrm{OKVS.Encode}(L), L = \left\{(x^{i}_{q}, \bigoplus_{j=v+1}^{n} F_{k_{i}^{j}}(x_{q}^{i})) \mid x^{i}_{q} \in X^{i}\right\}.$}\\
    \vspace{0.1cm}
    \hspace{-0.4cm}\resizebox{0.3\linewidth}{!}{
    $\mathcal{T}= \{T_{1},\dots T_{v-1}\}$.}
    \item Party $P_i$ in group A sends the $T_i$ to the central coordinator $P_v$ and all parties exchange the $(root_{X}, \Pi_{X}).$
\end{enumerate}

\paragraph{\underline{$\mathbf{Interaction}$ \((1^{\lambda}, \mathrm{st}, \mathcal{T}) \to \mathcal{R}\)}}
\begin{enumerate}
\item All Parties $P_i$ processes the verify procedure as:
\[
\mathcal{C}^j_{i} \leftarrow \mathrm{Merkle.Verify}(root_X^j, \Pi_X^j).
\]
Among all the output of verify procedure, if there is an $\perp$, output $\mathcal{R}= \perp$ and abort the procedure.
\item The central coordinator computes the key value set based on the $T_i$ from the parties in group A:
\[
A^{v} = \left\{ ( x_{q}^{v}, \bigoplus_{i=1}^{v-1} \text{OKVS.Decode}(T_{i}, x_{q}^{v}) ) \right\}_{q \in [n_l]}
\]
\item Parties in group B generate their key value set through the keys they received which was generated by the parties in Group A:
\[
A^{i} = \left\{ ( x_{q}^{i}, \bigoplus_{j=1}^{v-1} F_{k^{i}_j}(x_{q}^{i})) \right\}_{q \in [n_l]}
\]
$\mathcal{R}=\{A^{v},A^{v+1},\dots,A^{n}\}$.
\end{enumerate}

\paragraph{\underline{$\mathbf{Reconstruct}$ \((\mathrm{st},  \mathcal{R}) \to \mathcal{I}\)}}
\begin{enumerate}
    \item Party $P_i$ in $P_v,\dots,P_n$ invokes Zero-Sharing protocol on $x^{i}_j$ and obtains its share $S(K_i,x^{i}_j)$ for every $j \in [n_l]$.
    \item Party $P_{i\in [v,n-1]}$ and $P_n$ jointly invoke OPPRF according to their key value set $A^{i}$ as follows:
\begin{itemize}[leftmargin=0pt,labelsep=0.6em,nosep]
\item $P_v$ and $P_{v+1},\dots,P_{n-1}$ act as the sender, programming:
\vspace{-0.35cm}
\begin{adjustwidth}{-1.6cm}{0pt}
{\small\begin{align*}
    \mathcal{P}_v &= \left\{( x_{q}^{v}, S(K_v,x^{v}_q) \oplus \bigoplus_{i=1}^{v-1} \text{OKVS.Decode}(T_{i}, x_{q}^{v}) )\right\}_{q\in [n_l]}\\
    \mathcal{P}_i &= \left\{( x_{q}^{i}, S(K_i,x^{i}_q) \oplus \bigoplus_{j=1}^{v-1} F_{k^{i}_j}(x_{q}^{i}))\right\}_{q\in [n_l]}
\end{align*}}    
\end{adjustwidth}
    \item $P_n$ acts as the receiver with queries $\{x_{q}^{n}\}_{q\in [n_l]}b$ and obtains $\{(x_{q}^{n}, z_{q}^{i})\}_{q\in [n_l], i\in [v, n-1]}$ where $z_{q}^{i}$ is equal to the corresponding value if $x_{q}^{n}$ is included in in $\mathcal{P}_{v},\dots,\mathcal{P}_{n-1}$ and a pseudorandom value otherwise.
    \end{itemize}
    \item Party $P_n$ obtains the intersection \[\mathcal{I}=\left\{x^{n}_{q} \mid S(K_n, x^{n}_q) \oplus \bigoplus_{j=1}^{v-1} F_{k^{n}_j}(x_{q}^{n}) = \bigoplus_{i=v}^{n-1} z_{q}^{i}\right\}\]
\end{enumerate}
\end{tcolorbox}

\subsection{Proof of Two-party PSI}
\begin{theorem}\label{proof_Authenticated_PSI_Correctness}
$\mathbf{Correctness\,of\, Construction\,1:}$
First, we are going to analyze the correctness of this protocol. Before doing so, we need to examine the construction of the OKVS proposed in~\cite{RR22}. Let us define $\mathrm{row}(k_i) = \mathrm{row'}(k_i)||\hat{\mathrm{row}}(k_i)$, where $\mathrm{row'}(k_i)\in \{0, 1\}^{m'}$ is a uniformly random weight $ \omega $ vector, and $\hat{\mathrm{row}}(k_i) \in \mathbb{F}^{\hat{m}}$ for $\hat{m} = m - m'.$ where $m' \approx 1.23n_x$, The $\mathrm{OKVS.Encode}(x, \mathrm{H}^{\mathbb{B}}(x))$ algorithm of \cite{RR22} works as follows:
\[
\begin{bmatrix}
\mathrm{row}(x_1) \\
\vdots \\
\mathrm{row}(x_{n_x})
\end{bmatrix}\mathbf{P}^\top
=
\left(
\mathrm{H}^{\mathbb{B}}(x_1), \ldots, \mathrm{H}^{\mathbb{B}}(x_{n_x})
\right)^\top
\]
The \cite{RR22} uses the Triangulate and Back-substitution to obtain the $\mathbf{P}$ satisfies the above condition. The $\mathrm{OKVS.Decode}(\mathbf{P}, x)$ will return $\langle \mathbf{P},\mathrm{row}(x) \rangle$.

This OKVS follows the linearity for $x \in X$ according to the property of inner product:
\begingroup
\setlength{\mathindent}{0pt}
{\small\begin{align*}
\mathbf{C}' - \mathbf{B}' &= \Delta \cdot \mathbf{P} \\
\mathrm{Decode}(\mathbf{C}', x) - \mathrm{Decode}(\mathbf{B}', x) &= \Delta \cdot \mathrm{Decode}(\mathbf{P}, x) = \Delta \cdot \mathrm{H}^{\mathbb{B}}(x) \\
\mathrm{Decode}(\mathbf{C}', x) &= \mathrm{Decode}(\mathbf{B}', x) + \Delta \cdot \mathrm{H}^{\mathbb{B}}(x)
\end{align*}}
\endgroup
For $\mathcal{R} = \{\mathrm{H}^\circ\big(\mathrm{OKVS.Decode}(\mathbf{B'}, y) - \Delta \cdot \mathrm{H}^\mathbb{B}(y)\big) \;|\; y \in Y\},$ \\if $x \in X \bigcap Y:$
\begingroup
\setlength{\mathindent}{0pt}
\begin{align*}
\mathrm{Decode}(\mathbf{B}', x) -& \Delta \cdot \mathrm{H}^{\mathbb{B}}(x) \\
&= \langle \mathbf{B}', \mathrm{row}(x) \rangle - \Delta \cdot \mathrm{H}^{\mathbb{B}}(x)\\
&= \langle \mathbf{B} + \mathbf{A}' \cdot \Delta, \mathrm{row}(x) \rangle - \Delta \cdot \mathrm{H}^{\mathbb{B}}(x) \\
&= \langle \mathbf{B} + (\mathbf{A} + \mathbf{P}) \cdot \Delta, \mathrm{row}(x) \rangle - \Delta \cdot \mathrm{H}^{\mathbb{B}}(x) \\
&= \langle \mathbf{C} + \mathbf{P} \cdot \Delta, \mathrm{row}(x) \rangle - \Delta \cdot \mathrm{H}^{\mathbb{B}}(x) \\
&= \langle \mathbf{C}, \mathrm{row}(x) \rangle + \Delta \cdot \langle \mathbf{P}, \mathrm{row}(x) \rangle - \Delta \cdot \mathrm{H}^{\mathbb{B}}(x) \\
&= \langle \mathbf{C}, \mathrm{row}(x) \rangle + \Delta \mathrm{H}^{\mathbb{B}}(x) - \Delta \mathrm{H}^{\mathbb{B}}(x) \\
&= \mathrm{Decode}(\mathbf{C}, x).
\end{align*}
\endgroup
Regarding the parameters $\textit{out}$ and $\mathbb{B}$, the probability that an element $x \notin Y$ still satisfies the given equation is $2^{-\textit{out}} n_y = 2^{-\textit{out}+\log_{2}(n_y)}$. Consequently, the overall probability of a collision is $n_x 2^{-\textit{out}+\log_{2}(n_y)} = 2^{-\textit{out}+\log_{2}(n_y n_x)} = 2^{-\lambda}$. To achieve the desired functionality against a semi-honest adversary, the size of $\mathbb{B}$ must be chosen such that $\lvert \mathbb{B} \rvert \geq 2^{\lambda + \log_{2}(n_x) + \log_{2}(n_y)}$, ensuring that the collision probability constraint $n_x n_y \lvert \mathbb{B} \rvert^{-1} \leq 2^{-\lambda}$ is satisfied. 

Thus, the Receiver only need to compare the $\mathcal{R}$ and $\mathcal{R}'$, then can obtain the intersection $X \bigcap Y$, and the Correctness of the protocol can be obtained according to the property of building blocks (Merkle Tree, OKVS, VOLE) Definitions~\ref{MerkleTree_Correctness}, \ref{OKVScorrectness}, \ref{VOLE_Correctness} directly.    
\end{theorem}
\noindent Now, we turn to the proof of the protocol's Integrity and security. For the integrity of our protocol, we need to prove the property in Definition~\ref{AuthenticatedPSI_integrity}. 
\begin{theorem}\label{proof_Authenticated_PSI_Integrity}
$\mathbf{Integrity\,of\, Construction\,1:}$
We assume there is an adversarial participant attempting to tamper with input set to gain additional information, without lose of generality, we assume the original input set $X = \{x_1,\dots,x_{n_x}\}$ has been modified to $X^{*} = \{x_1,\dots,x^{*}_l,\dots,x_{n_x}\}$ for $x_i \in \{0,1\}^{*}$ where $x^{*}_l = x_l \oplus \Delta_{x}$, and the according tree path $\pi_{x^{*}_l} \leftarrow \mathrm{Merkle.GenPath}(1^{\lambda}, X^{*}, x^{*}_l)$ where $\pi_{x^{*}_l} = \pi_{x_l} \oplus \Delta_{\pi}$. We prove the integrity property by contradiction. The following holds:
    \[
    \Pr \left[
    \begin{array}{rl}
    & (\mathrm{st}_{\mathcal{A}}, \mathcal{T}) \leftarrow \mathcal{A}(1^{\lambda}, X^*) \\
    \mathcal{I} \neq \perp : & \mathcal{R} \leftarrow \mathrm{Interaction}(1^{\lambda}, \mathrm{st}_{\mathcal{A}}, \mathcal{T}) \\
    & \mathcal{I} \leftarrow \mathrm{Reconstruct}(\mathrm{st}_{\mathcal{A}}, \mathcal{R}) \\
    \end{array}
    \right] \geq \mathrm{v}
    \]
where $\mathrm{v}$ is non-negligible in the security parameter $\lambda$. We can rewrite the above probability in details as:\\
    \hspace{-0.5cm}\resizebox{0.5\textwidth}{!}{$\Pr \left[
     \begin{array}{c l}
     \mathcal{I} \neq \perp : & 
     \hspace{-0.5cm}\begin{array}{l}
     (\mathrm{st}, \mathcal{T}) \leftarrow \mathrm{Transform}(1^{\lambda}, X^*)\\
     \mathcal{R} \leftarrow \mathrm{Interaction}(1^{\lambda}, \mathrm{st}, \mathcal{T}):\\ 
     \quad\quad\, \{\pi_{i}\}_{i \neq l} \leftarrow \mathrm{Merkle.GenPath}(1^{\lambda}, X^{*}, \{x_{i}\}_{i \neq l})\\ 
     \quad\quad \pi_{l} \oplus \Delta_{\pi} \leftarrow \mathrm{Merkle.GenPath}(1^{\lambda}, X^{*}, x^{*}_{l})\\
     \quad\quad \Pi^{*} = \{\pi_1, \dots, \pi^{*}_{l}, \dots, \pi_{n_{x}}\}\\ 
     \mathcal{I} \leftarrow 
     \left\{
     \begin{aligned}
     & \perp && \text{if} \perp \leftarrow \mathrm{Merkle.Verify}(\text{root}, \Pi^{*})\\
     & \mathrm{Reconstruct}(\mathrm{st}, \mathcal{R}) && \text{otherwise}.
     \end{aligned}
     \right.
     \end{array}
     \end{array}
    \hspace{-0.45cm}\right] \geq \mathrm{v}$}\\
    
According to the Definition~\ref{MerkleTree_Correctness}, the output $\mathcal{I} \neq \perp$ only when the output of $\mathrm{Merkle.Verify}(\text{root}, \Pi^{*})$ is not $\perp$ which indicates $1 \leftarrow \mathrm{Merkle.Verify}(\text{root},X^{*}, x^{*}_{l})$. Considering $x^{*}_l = x_l \oplus \Delta_{x}$ and $\pi_{x^{*}_l} = \pi_{x_l} \oplus \Delta_{\pi}:$
\begin{itemize}
    \item if $\Delta_{x} \neq 0$, the adversary break the Soundness in Definition~\ref{MerkleTree_Soundness}, which means that the $x^{*}_l \neq x_l$ passes the verify procedure successfully.
    \item if $\Delta_{x} = 0$, but $\Delta_{\pi} \neq 0$, the adversary break the Uniqueness in Definition~\ref{MerkleTree_Uniqueness} which means the one $x$ passes two verify procedure under $\pi_{x_l}$ and $\pi^{*}_{x_l}$ and $\pi_{x_l} \neq \pi^{*}_{x_l}$.
\end{itemize}
Thus, the probability satisfies the probability $\geq \mathrm{v}$ only when $\Delta_{x} = 0$ and $\Delta_{\pi} = 0$ which means the input set has not been tempered, thus we proved that our construction satisfies the integrity of authenticated PSI in Definition~\ref{AuthenticatedPSI_integrity}.  
\end{theorem}
After the above content, we proceed to prove the security in Definition~\ref{AuthenticatedPSI_security} of our protocol, which is a crucial part of the security proof.
\begin{theorem}\label{proof_Authenticated_PSI_Security}
$\mathbf{Security\,of\, Construction\,1:}$
For our proctol, there is an efficient adversary $\mathcal{A} = (\mathcal{A}_0, \mathcal{A}_1)$, we assume the adversary tempered the input set to $X^{*} = \{x_1,\dots,x^{*}_l,\dots,x_{n_x}\}$ for $x_i \in \{0,1\}^{*}$ where $x^{*}_l = x_l \oplus \Delta_{x}$ and its tree path $\pi_{x^{*}_l} = \pi_{x_l} \oplus \Delta_{\pi}$, we can prase the real distribution of information as follows:\\

    \hspace{-0.5cm}\resizebox{0.5\textwidth}{!}{
    $\mathrm{REAL}_{\mathcal{A}, X, \lambda}= \left \{
     \begin{array}{c l}
     \hat{\mathcal{V}}: & 
     \begin{array}{l}
     (\mathrm{st}_\mathcal{A}, \mathcal{T}) \leftarrow \mathcal{A}_0(1^{\lambda}, X^*)\\
     \mathcal{R} \leftarrow \mathrm{Interaction}(1^{\lambda}, \mathrm{st}_\mathcal{A}, \mathcal{T}):\\ 
     \quad\quad\, \{\pi_{i}\}_{i \neq l} \leftarrow \mathrm{Merkle.GenPath}(1^{\lambda}, X^{*}, \{x_{i}\}_{i \neq l})\\ 
     \quad\quad \pi_{l} \oplus \Delta_{\pi} \leftarrow \mathrm{Merkle.GenPath}(1^{\lambda}, X^{*}, x^{*}_{l})\\
     \quad\quad \Pi^{*} = \{\pi_1, \dots, \pi^{*}_{l}, \dots, \pi_{n_{x}}\}\\ 
     \mathcal{I} \leftarrow 
     \left\{
     \begin{aligned}
     & \perp && \text{if } \perp \leftarrow \mathrm{Merkle.Verify}(\text{root}, \Pi^{*})\\
     & \mathrm{Reconstruct}(\mathrm{st}, \mathcal{R}) && \text{otherwise}.
     \end{aligned}
     \right.\\
       a \leftarrow \mathbbm{1}\{\mathcal{I} \neq \perp \}\\
      \hat{\mathcal{V}} \leftarrow \mathcal{A}_1(\mathrm{st}_{\mathcal{A}}, a)
     \end{array}
     \end{array}
    \hspace{-0.5cm}\right\}$}\\
    
We parse the ideal distribution with simulator $\mathcal{S}=(\mathcal{S}_0, \mathcal{S}_1)$ as follows:
     {\small\[
     \mathrm{IDEAL}_{\mathcal{A}, \mathcal{S}, X, \lambda} = \left\{
     \begin{array}{c l}
     \mathcal{V} : & 
     \begin{array}{l}
     (\mathrm{st}_{\mathcal{A}}, \mathcal{T}) \leftarrow \mathcal{A}_0(1^{\lambda}, X^*)\\
     (\mathrm{st}_{\mathcal{S}}, \mathcal{I}) \leftarrow \mathcal{S}_0(1^{\lambda}, \mathcal{T}) \\
     a \leftarrow \mathcal{S}_1(\mathrm{st}_{\mathcal{S}}, \mathcal{I}) \\
     \mathcal{V} \leftarrow \mathcal{A}_1(\mathrm{st}_{\mathcal{A}}, a)
     \end{array}
     \end{array}
     \right\}
     \]}
For any adversary $\mathcal{A} = (\mathcal{A}_0, \mathcal{A}_1)$, the simulator $\mathcal{S}=(\mathcal{S}_0, \mathcal{S}_1)$ works as follows where the $\mathcal{S}_{\mathrm{PSI}}$ is the simulator mentioned in standard PSI in Definition~\ref{PSI_Security} which is used to sample the interaction element from the information space:
\[
\begin{array}{|c|c|}
\hline
\textbf{Simulator}\: S_0 (1^\lambda, \mathcal{T}) & \textbf{Simulator}\: S_1 (st_S, \mathcal{I}) \\
\hline
\begin{array}{rl}
1: & \mathcal{I} \leftarrow \mathcal{S}_{\mathrm{PSI}}(1^{\lambda}, \{0,1\}^{*}) \\
2: & (x^{*}_{l}, \pi^{*}_l) \leftarrow \mathcal{T}\\
3: & st_S \leftarrow (x^{*}_{l}, \pi^{*}_l) \\
4: & \textbf{return} (st_S, \mathcal{I})
\end{array}
&
\begin{array}{rl}
1: & (x^{*}_{l}, \pi^{*}_l) \leftarrow st_S \\
2: & \Delta_x \leftarrow x^{*}_{l} \oplus x_{l} \\
3: & \Delta_{\pi} \leftarrow \pi^{*}_{l} \oplus \pi_{l}\\
4: & a \leftarrow \mathbbm{1}\{\Delta = 0\} \\
5: & \textbf{return } a
\end{array}
\\
\hline
\end{array}
\]
We proceed to demonstrate that the real and ideal distributions are computationally indistinguishable, thereby establishing that the scheme described in \hyperref[construction1]{Construction 1} follows the authenticated PSI security in Definition~\ref{AuthenticatedPSI_security}. We first construct four hybrid distribution: $\mathrm{H}_{0}, \mathrm{H}_{1}, \mathrm{H}_{2}, \mathrm{H}_{3}$:
\begin{itemize}
    \item $\mathbf{H_0:}$ The real distribution $\mathrm{REAL}_{\mathcal{A},X,\lambda}$ where $x^{*}_l = x_l \oplus \Delta_{x}$ and its tree path $\pi_{x^{*}_l} = \pi_{x_l} \oplus \Delta_{\pi}$.
    \item $\mathbf{H_1:}$ Same as the $\mathbf{H_0}$, we change the conditional statemrnt of bit $a$, instead of $\mathcal{I} \neq \perp$:\\
    
    \hspace{-0.8cm}\resizebox{0.5\textwidth}{!}{
    $\mathrm{H}_1 = \left \{
     \begin{array}{c l}
     \hat{\mathcal{V}}: & 
     \begin{array}{l}
     (\mathrm{st}_\mathcal{A}, \mathcal{T}) \leftarrow \mathcal{A}_0(1^{\lambda}, X^*)\\
     \mathcal{R} \leftarrow \mathrm{Interaction}(1^{\lambda}, \mathrm{st}_\mathcal{A}, \mathcal{T}):\\ 
     \quad\quad\, \{\pi_{i}\}_{i \neq l} \leftarrow \mathrm{Merkle.GenPath}(1^{\lambda}, X^{*}, \{x_{i}\}_{i \neq l})\\ 
     \quad\quad \pi_{l} \oplus \Delta_{\pi} \leftarrow \mathrm{Merkle.GenPath}(1^{\lambda}, X^{*}, x^{*}_{l})\\
     \quad\quad \Pi^{*} = \{\pi_1, \dots, \pi^{*}_{l}, \dots, \pi_{n_{x}}\}\\ 
     \mathcal{I} \leftarrow 
     \left\{
     \begin{aligned}
     & \perp && \text{if} \perp \leftarrow \mathrm{Merkle.Verify}(\text{root}, \Pi^{*})\\
     & \mathrm{Reconstruct}(\mathrm{st}, \mathcal{R}) && \text{otherwise}.
     \end{aligned}
     \right.\\
       a \leftarrow \mathbbm{1}\{\Delta_x = 0 \text{ and }\Delta_{\pi} = 0  \}\\
      \hat{\mathcal{V}} \leftarrow \mathcal{A}_1(\mathrm{st}_{\mathcal{A}}, a)
     \end{array}
     \end{array}
    \hspace{-0.4cm}\right\}$}\\
    
    \item $\mathbf{H_2:}$ For this hybrid distribution, comparing to the hybrid $\mathbf{H_1}$, we consider about use the simulator $\mathcal{S}_{\mathrm{PSI}}$ to directly sample the intersection elements from the data space to simulate the generation of the intersection:\\
    
    \hspace{-0.8cm}\resizebox{0.5\textwidth}{!}{
    $\mathrm{H}_2 = \left \{
     \begin{array}{c l}
     \hat{\mathcal{V}}: & 
     \begin{array}{l}
     (\mathrm{st}_\mathcal{A}, \mathcal{T}) \leftarrow \mathcal{A}_0(1^{\lambda}, X^*)\\
     \mathcal{R} \leftarrow \mathrm{Interaction}(1^{\lambda}, \mathrm{st}_\mathcal{A}, \mathcal{T}):\\ 
     \quad\quad\, \{\pi_{i}\}_{i \neq l} \leftarrow \mathrm{Merkle.GenPath}(1^{\lambda}, X^{*}, \{x_{i}\}_{i \neq l})\\ 
     \quad\quad \pi_{l} \oplus \Delta_{\pi} \leftarrow \mathrm{Merkle.GenPath}(1^{\lambda}, X^{*}, x^{*}_{l})\\
     \quad\quad \Pi^{*} = \{\pi_1, \dots, \pi^{*}_{l}, \dots, \pi_{n_{x}}\}\\ 
     \mathcal{I} \leftarrow \mathcal{S}_{\mathrm{PSI}}(1^{\lambda}, \{0,1\}^{*})\\
       a \leftarrow \mathbbm{1}\{\Delta_x = 0 \text{ and }\Delta_{\pi} = 0  \}\\
      \hat{\mathcal{V}} \leftarrow \mathcal{A}_1(\mathrm{st}_{\mathcal{A}}, a)
     \end{array}
     \end{array}
    \hspace{-0.4cm}\right\}$}\\

    \item $\mathbf{H_3:}$ The ideal distribution $\mathrm{IDEAL}_{\mathcal{A}, \mathcal{S}, X, \lambda}$.
\end{itemize}
We now tend to demonstrate that every pair of consecutive hybrids is indistinguishable and use $\mathrm{E}_i$ for $i \in \{0,1,2,3\}$ to denote the event of distribution:
\begin{itemize}
    \item $\mathbf{\mathrm{H}_0 \to \mathrm{H}_1:}$ $\mathrm{H}_1$ is identical to $\mathrm{H}_0$ except for the acceptance of the bit $a$. When comparing the conditions $\mathcal{I} \neq \perp$ and $\Delta_x = 0, \Delta_\pi = 0$, based on Theorem~\ref{proof_Authenticated_PSI_Integrity} and Definition~\ref{MerkleTree_Correctness}, we know that:
\[
\Pr[\mathcal{I} \neq \perp \mid \Delta \neq 0] \leq \mathrm{negl}(\lambda).
\]
When $\Delta = 0$, the probability of $\mathcal{I} \neq \perp$ should approach 1, i.e.,
\[
\Pr[\mathcal{I} \neq \perp \mid \Delta = 0] \approx 1.
\]
Thus, the overall probability $\Pr[\mathcal{I} \neq \perp]$ can be expressed as:
\begin{align*}
\Pr[\mathcal{I} \neq \perp] & = \Pr[\mathcal{I} \neq \perp \mid \Delta = 0] \cdot \Pr[\Delta = 0] \\&+ \Pr[\mathcal{I} \neq \perp \mid \Delta \neq 0] \cdot \Pr[\Delta \neq 0].  
\end{align*}

Since $\Pr[\mathcal{I} \neq \perp \mid \Delta = 0] \approx 1$ and $\Pr[\mathcal{I} \neq \perp \mid \Delta \neq 0]$ is negligible, we have:
\[
\Pr[\mathcal{I} \neq \perp] \approx \Pr[\Delta = 0].
\]
Additionally, the difference between two probabilities can be bounded as:
\begingroup
\setlength{\mathindent}{0pt}
\begin{align*}
   | \Pr[a \leftarrow \mathbbm{1}\{\Delta_x = 0 & \wedge  \Delta_\pi = 0\}] \\ &- \Pr[a \leftarrow \mathbbm{1}\{\mathcal{I} \neq \perp\}] | \leq \mathrm{negl}(\lambda). 
\end{align*}
\endgroup
This implies that:
\[
\left| \Pr[\mathrm{E}_1] - \Pr[\mathrm{E}_0] \right| \leq \mathrm{negl}(\lambda).
\]
    \item $\mathbf{\mathrm{H}_1 \to \mathrm{H}_2:}$ The only difference between hybrids $\mathrm{H}_1$ and $\mathrm{H}_2$ is the generation of intersection information $\mathcal{I},$ according to the security of standard PSI in Definition~\ref{PSI_Security}, we have:
    \[
    \left| \Pr[\mathrm{E}_2] - \Pr[\mathrm{E}_1] \right| \leq \mathrm{negl}(\lambda).
    \]
    \item $\mathbf{\mathrm{H}_2 \to \mathrm{H}_3:}$ Actually, the hybrid $\mathrm{H}_3$ can be seen as a reformulation of $\mathrm{H}_2$, achieved by simply omitting the results of the interaction phase. In the original protocol, $\mathcal{R}$ played a critical role in the third phase, Reconstruct, to derive the final intersection $\mathcal{I}$. However, with the introduction of the simulator and the new method for generating $\mathcal{I}$, the focus of the second phase has shifted primarily to verifying the Merkle tree paths of the input data. Consequently, we have:
    \[
    \left| \Pr[\mathrm{E}_3] - \Pr[\mathrm{E}_2] \right| = 0.
    \]
    By a series of indistinguishable comparisons, we finally obtain that:
    \[
    \mathrm{REAL}_{\mathcal{A},X,\lambda} = \mathrm{H}_0 \approx_c \mathrm{H}_1 \approx_c \mathrm{H}_2 = \mathrm{H}_3 = \mathrm{IDEAL}_{\mathcal{A}, \mathcal{S}, X, \lambda}.
    \]
\end{itemize}
\end{theorem}

\subsection{Proof of Multi-party PSI}

\begin{theorem}\label{proof_Authenticated_PSI_Correctness2}
$\mathbf{Correctness\,of\, Construction\,2:}$We now tend to present the correctness of \hyperref[construction2]{Construction 2}, if a value $x$ is included in the intersection of all parties such that $x \in \bigcap_{i\in [n]}X^{i}:$
\begingroup
\setlength{\mathindent}{0pt}
{\small \begin{align*}
&S(K_n, x) 
 \oplus  \bigoplus_{j=1}^{v-1} F_{k^{n}_j}(x) 
\oplus  \bigoplus_{i=v}^{n-1} z_{q}^{i}  \\
&= S(K_n, x) 
\oplus \bigoplus_{j=1}^{v-1} F_{k^{n}_j}(x)  
\oplus \Big( S(K_v, x) 
\oplus \bigoplus_{i=1}^{v-1} \text{OKVS.Decode}(T_{i}, x) \Big) \\
&\quad \oplus \Big( \bigoplus_{i=v+1}^{n-1} S(k_{i}, x) 
\oplus \bigoplus_{j=1}^{v-1} F_{k^{i}_j}(x) \Big) \\
&= S(K_n, x) \oplus S(K_v, x) 
\oplus \bigoplus_{i=v+1}^{n-1} S(k_{i}, x) 
\oplus \bigoplus_{j=1}^{v-1} F_{k^{n}_j}(x) \\
&\quad \oplus \bigoplus_{i=1}^{v-1} \bigoplus_{j=v+1}^{n} F_{k^{j}_i}(x) 
\oplus \bigoplus_{i=v+1}^{n-1} \bigoplus_{j=1}^{v-1} F_{k^{i}_j}(x) \\
&= \Big(\bigoplus_{i=v}^{n} S(k_{i}, x) \Big)
\oplus \Big( \bigoplus_{i=1}^{v-1} \bigoplus_{j=v+1}^{n} F_{k^{j}_i}(x) \Big)
\oplus \Big( \bigoplus_{i=v+1}^{n} \bigoplus_{j=1}^{v-1} F_{k^{i}_j}(x) \Big)
\end{align*}}
\endgroup
We observe that the parties \( P_v, \dots, P_n \) invoke the Zero-Sharing protocol. Consequently, we have $\bigoplus_{i=v}^{n} S(k_{i}, x) = 0$ according to the Zero-Sharing process mentioned above. For the remaining part of the equation $\left(\bigoplus_{i=1}^{v-1} \bigoplus_{j=v+1}^{n} F_{k^{j}_i}(x)\right) \oplus \left(\bigoplus_{i=v+1}^{n} \bigoplus_{j=1}^{v-1} F_{k^{i}_j}(x)\right)$, in the left part, the keys used by OPPRF are come from \( P_1, \dots, P_{v-1} \), which are generated for \( P_{v+1}, \dots, P_n \). Similarly, in the right part, the same keys are used; however, the users switch from members of Group A to members of Group B. This is equivalent to each key being used twice with the same \( x \), resulting in the final XOR sum being zero. Therefore, the above content indicates that the output \( \mathcal{I} \) that satisfies the conditions is the intersection set of all participants.
\end{theorem}
As for the integrity and security of the construction, they heavily rely on proofs of the relevant properties of Merkle trees. The specific process is consistent with the proof of \hyperref[construction1]{Construction 1} in Theorem~\ref{proof_Authenticated_PSI_Integrity},\ref{proof_Authenticated_PSI_Security}.
\section{Implementation and Evaluation}\label{sec5}
\subsection{Implementation Details and Performance Results}
\begin{table*}[ht]\caption{Performance of Merkle Tree-volePSI}
\label{tab1}
\centering
\resizebox{0.7\textwidth}{!}{%
    \begin{tabular}{l|llll|lccc}
    \hline
    \multicolumn{1}{c|}{\multirow{2}{*}{\textbf{Protocol}}} &
      \multicolumn{4}{c|}{\textbf{Times(ms)}} &
      \multicolumn{4}{c}{\textbf{Communication(bits/n)}} \\ \cline{2-9} 
    \multicolumn{1}{c|}{} &
      \multicolumn{1}{c}{$2^{10}$} &
      \multicolumn{1}{c}{$2^{12}$} &
      \multicolumn{1}{c}{$2^{14}$} &
      \multicolumn{1}{c|}{$2^{16}$} &
      \multicolumn{1}{c}{$2^{10}$} &
      $2^{12}$ &
      $2^{14}$ &
      $2^{16}$ \\ \hline
    Merklecpp+volePSI(useQC) &
      49.67 &
      123.57 &
      509.39 &
      2050.78 &
       &
      & &
      \\
    Merklecpp+volePSI(useSliver) &
      39.73 &
      124.49 &
      479.52 &
      1962.60 &
      462$n$ &
      437$n$ &
      455$n$ &
      467$n$ \\
    Merklecpp+volePSI(useSliver,MT) &
      36.83 &
      94.31 &
      460.33 &
      1897.89 &
       &
      & &
      \\ \hline
    Merklecpp+volePSI(malicious) &
      49.84 &
      134.35 &
      512.95 &
      2160.41 &
      1257$n$ &
      1187$n$ &
      1213$n$ &
      1263$n$ \\ \hline
    \end{tabular}
}
\end{table*}
\begin{table*}[ht]\caption{Performance of Merkle Tree-mPSI \quad \footnotesize
    \textbf{Notes:} times in milliseconds (ms). \(t\) = number of corrupted parties.
  }
\label{tab2}
\centering
\resizebox{0.7\textwidth}{!}{%
    \begin{tabular}{l|l|lllclcl}
    \hline
    \multicolumn{1}{c|}{\multirow{2}{*}{\textbf{Protocol}}} &
      \multicolumn{1}{c|}{\textbf{$n$}} &
      \multicolumn{1}{c}{3} &
      \multicolumn{2}{c}{4} &
      \multicolumn{2}{c}{5} &
      \multicolumn{2}{c}{8} \\ \cline{2-9} 
    \multicolumn{1}{c|}{} &
      \diagbox[height=0.7cm,width=1.2cm]{$n_l$}{$t$} &
      \multicolumn{1}{c}{1} &
      \multicolumn{1}{c}{1} &
      \multicolumn{1}{c}{2} &
      1 &
      \multicolumn{1}{c}{3} &
      1 &
      \multicolumn{1}{c}{4} \\ \hline
               & $2^{8}$  & 117.13 & 118.21 & 183.77 & \multicolumn{1}{l}{116.45} & 202.76 & \multicolumn{1}{l}{134.21} & 205.76 \\
    Merklecpp+mPSI & $2^{10}$ & 165.13 & 186.54 & 241.25 & 188.76 & \multicolumn{1}{c}{261.90} & 183.03 & \multicolumn{1}{c}{274.56} \\ 
               & $2^{12}$ & 791.61 & 795.47 & 910.91 & \multicolumn{1}{l}{817.34} & 907.18 & \multicolumn{1}{l}{818.51} & 935.47 \\ \cline{2-9} \hline
    \end{tabular}
} 
\end{table*}
Our construction aims to enhance the security of traditional PSI schemes, particularly emphasizing the integrity property required by authenticated PSI. As an extension of traditional PSI functionality, authenticated PSI introduces additional security guarantees. To achieve authenticated PSI, auxiliary tools such as Merkle trees or other potential solutions (e.g., digital signatures and secret sharing) are required. The introduction of these new tools and schemes impacts the overall performance of PSI. Furthermore, the efficiency of the entire scheme is highly dependent on the performance of the underlying PSI scheme, which is why \hyperref[construction1]{Construction 1} and \hyperref[construction2]{Construction 2} in this paper integrate one of the current best two-party PSI schemes and one of the best multi-party PSI schemes, respectively. Even in the absence of comparable PSI schemes within the same category, the implementation of our scheme partially demonstrates its feasibility, and related issues can be further optimized in future work.

Our scheme is entirely implemented in C++ and used 3x NVIDIA A40 48GB GPU and 256GB of RAM. In both constructions, the Merkle Tree component relies on Microsoft's \href{https://github.com/microsoft/merklecpp}{merklecpp} project. Additionally, the underlying two-party PSI scheme is based on \cite{RR22} and its implementation \href{https://github.com/Visa-Research/volepsi}{github.com/Visa-Research/volepsi}, which optimizes the VOLE and OKVS components from \cite{RS21} and incorporates the subfield-VOLE optimizations from \cite{CRR21}. For the OKVS component, it replace the original PaXos method from \cite{PRTY20} with improvements from \cite{GPR21} and their implementation \href{https://github.com/cryptobiu/OBDBasedPSI}{github.com/cryptobiu/OBDBasedPSI}, modifying the row weight from $\omega=2$ in \cite{PRTY20} to $\omega=3$ and performing targeted optimizations. Furthermore, the scheme utilizes Oblivious Transfers from \href{https://github.com/osu-crypto/libOTe}{libOTe}. The multi-party PSI scheme relies on \cite{NTY21} and its implementation \href{https://github.com/asu-crypto/mPSI}{github.com/asu-crypto/mPSI}. The OPRF and table-based OPPRF constructions are derived from \cite{KKRT16} and \cite{KMP+17}, respectively, while the OKVS construction is based on \cite{PRTY20}. The usage of pseudorandom functions (PRF) depends on the AES-NI instruction set. Additionally, this scheme also employs OT from \href{https://github.com/osu-crypto/libOTe}{libOTe}. Both constructions adopt a statistical security parameter $\lambda = 40$ and a computational security parameter $\kappa = 128$.

In Table \ref{tab1}, \textbf{useQC} represents the utilization of the QuasiCyclic VOLE encoder as described in \cite{BCG+19}, while \textbf{useSilver} highlights the adoption of the Silver VOLE encoder from \cite{CRR21}. The use of these VOLE encoders significantly improves the speed of the PSI component. Aligning with the description in \cite{RR22}, "MT" indicates that each party uses 4 threads to execute the process. In Table \ref{tab2}, the configurations $(n, t) \in \{(3,1), (4,\{1,2\}), (5,\{1,3\}), (8,\{1,4\}$ $)\}$ are compared to evaluate the efficiency under different levels of collusion. Additionally, due to the construction employed in the scheme of \cite{NTY21}, which is based on \cite{CDG21}, the number of collusions is restricted to $t \leq \frac{n}{2}$.

For our proposed scheme, the most time-consuming component lies in the merkle tree operations. Specifically, the time required for generating the tree root and performing insertions is minimal, typically within the range of 10 milliseconds. The major computational overhead stems from the verification of merkle tree paths. This is currently a key bottleneck of authenticated PSI, as achieving data integrity comes at a significant cost. With increasing data set sizes, the depth of the merkle tree grows correspondingly, leading to a substantial increase in communication overhead and the time required by participants to verify integrity. However, when dealing with smaller datasets (e.g., less than $2^{10}$ elements), the impact of merkle tree operations becomes negligible. This aligns with the intended application scenarios of authenticated PSI, which primarily focus on managing smaller-scale datasets while ensuring data integrity.

\subsection{Effectiveness of Small-Domain Integrity Attacks}
\begin{figure*}[!t]
  \centering
  \centering\subfloat[\small{Reconstruction vs.\ probe budget (log scale).}]{
    \includegraphics[width=0.485\textwidth]{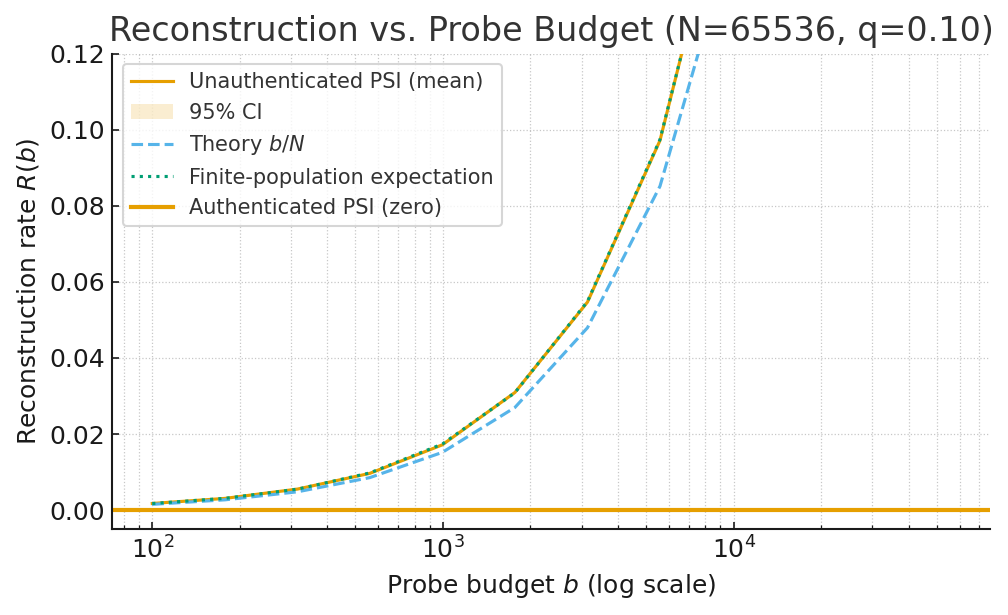}
  }
  \hfil
  \centering\subfloat[\small{Gain--loss tradeoff with fixed-cardinality substitution ($b=k$).}]{
    \includegraphics[width=0.485\textwidth]{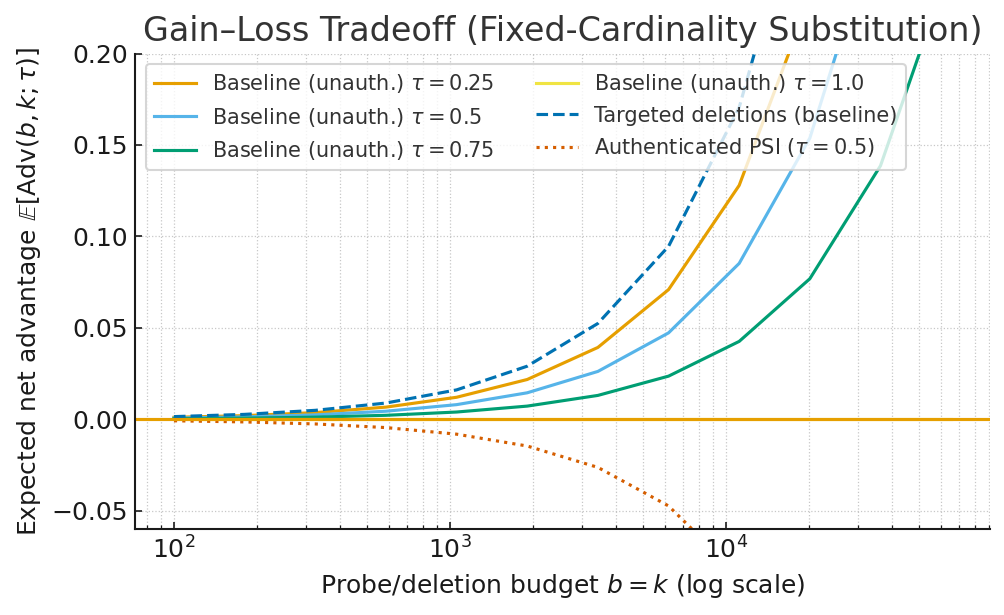}
  }
  \caption{Attack effectiveness and utility for small-domain probing. 
  Left: leakage grows with probe budget while authenticated PSI remains at zero. 
  Right: the net advantage of substitution depends on deletion penalty $\tau$.}
  \vspace{-0.6em}
  \label{fig3}
\end{figure*}
To highlight the practical advantage of authenticated PSI, we adopt a targeted threat model as follows and Figure~\ref{fig3} instantiates this model and reports reconstruction and the gain–loss tradeoff, showing that authenticated PSI blocks probe-induced leakage and removes the attacker’s net advantage.

\textbf{Threat model.}
We consider an \emph{honest-but-inconsistent} adversary that follows the message flow but submits a changed set
\(S'_A=(S_A\setminus D)\cup P\), where \(D\subseteq S_A\) are deletions and \(P\subseteq\mathcal{D}\setminus S_A\) are
\emph{probe elements} used solely to test membership. Let \(k:=|D|\) and \(b:=|P|\); a stealthy attacker may keep the reported
cardinality unchanged by choosing \(k=b\) (fixed-cardinality probing). The domain is enumerable \(\mathcal{D}\) with size
\(N\); the victim holds \(S_H\subseteq\mathcal{D}\) with \(|S_H|=s\) and prevalence \(q=s/N\).

\textbf{Metrics.}
We report (i) \emph{probe-only reconstruction} \(R_P(b)=|S_H\cap P|/|S_H|\), which isolates leakage caused by probes;
(ii) \emph{leakage per probe} \(L(b)=|S_H\cap P|/|P|\); and for completeness we also consider the \emph{deletion loss}
\(\Delta_{\mathrm{del}}(k)=|S_H\cap D|/|S_H|\), since \(S'_A\) removes those potential true matches. We summarize the attacker’s gain–loss tradeoff by
\[
\mathsf{Adv}(b,k;\tau)\ :=\ R_P(b)\ -\ \tau\,\Delta_{\mathrm{del}}(k),
\]
where $\tau\!\ge\!0$ weighs the value of kept matches relative to new information (larger $\tau$ penalizes deletions more).

\textbf{Baseline (unauthenticated PSI).}
Under random probing each probe hits with probability \(q\), so \(\mathbb{E}[R_P(b)]=b/N\) (saturating at \(1\) when \(b\ge N\)), $\mathbb{E}[\Delta_{\mathrm{del}}(k)]=k/N$ and $\mathbb{E}[\mathsf{Adv}]=\frac{b-\tau k}{N};
\quad\text{for fixed-cardinality }k=b,\ \mathsf{Adv}>0\ \Leftrightarrow\ \tau<1.$ (With targeted deletions $\Delta_{\mathrm{del}}\!\approx\!0$, so $\mathsf{Adv}\!\approx\!R_P(b)$.)
Authenticated PSI enforces $R_P(b)=0$, yielding $\mathsf{Adv}\le 0$ for any $\tau>0$.

\textbf{Effect of authenticated PSI.}
Every submitted element must carry a valid Merkle-based membership witness against the committed root \(C_A\); probe elements
not in the commitment have no witness and are rejected. Consequently \(R_P(b)=0\), \(L(b)=0\) within this
threat model.

\section{Conclusion}\label{sec6}
In this work, we introduce a novel definition of integrity for PSI and further propose authenticated PSI as a means of addressing data integrity attacks. By integrating Merkle Trees \cite{Mer89} with state-of-the-art two-party PSI protocols \cite{RS21,RR22} and multi-party PSI protocols \cite{NTY21}, we construct both two-party and multi-party authenticated PSI schemes and demonstrate their implementation and performance. Authenticated PSI augments the strong privacy guarantees of traditional PSI with robust data authentication was shown under our threat model evaluation.

 In addition, the specific type of authenticated PSI closely depends on the underlying PSI scheme (whether honest-majority or malicious, two-party or multi-party, balanced or unbalanced). Consequently, advancements in base PSI constructions will naturally drive improvements in authenticated PSI. With regard to the verification block, the sequential leaf-computation algorithm and authentication-path generation for Merkle trees have essentially reached optimal complexity \cite{Szy04}. Nonetheless, alternative methods such as digital signatures \cite{LZG+24} or function secret sharing could also be employed to realize integrity verification, is it feasible to design multi-party authenticated PSI protocols that matches the best standard schemes? As these building blocks continue to mature, authenticated PSI will play a pivotal role in supply-chain management, cross-organizational collaboration, as well as in medical data sharing and collaborative diagnostics.

\section{Acknowledgements}
This work was supported by Information Security School-Enterprise
Joint Laboratory (Dongguan Institute for Advanced Study, Greater Bay
Area) and Project (No. H24120002, H24120003) and High-level Talent Research Start-up Project Funding of Henan Academy of Sciences (Project NO. 232019\\007), and in part by Joint Fund of Henan Province Science and Technology R$\&$D Program (Project NO. 225200810036).

\end{document}